\documentclass[aps,prd,amsmath,amssymb,amsfonts,showpacs,superscriptaddress,floatfix,twocolumn,nofootinbib,longbibliography]{revtex4-1}
\usepackage{amsmath}
\usepackage{amssymb}
\usepackage{graphicx}
\usepackage{bbm}
\usepackage{caption}
\usepackage{subcaption}
\usepackage{epsfig}
\usepackage{braket}
\usepackage{slashed}
\usepackage{bm}
\usepackage[usenames,dvipsnames]{color}
\usepackage{color}
\usepackage{float}
\usepackage{hyperref}

\def\be{\begin{equation}}
\def\ee{\end{equation}}

\def\bea{\begin{eqnarray}}
\def\eea{\end{eqnarray}}
\begin{document}

\title{Captain Einstein: a VR experience of relativity} 

\author{Karel Van Acoleyen}
\affiliation{Department of Physics and Astronomy,
Ghent
 University,
  Krijgslaan 281, S9, 9000 Gent, Belgium}

\author{Jos Van Doorsselaere}
  \affiliation{Department of Physics and Astronomy,
Ghent
 University,
  Krijgslaan 281, S9, 9000 Gent, Belgium}

\begin{abstract}
\noindent Captain Einstein is a virtual reality (VR) movie that takes you on a boat trip in a world with a slow speed of light. This allows for a direct experience of the theory of special relativity, much in the same spirit as in the Mr. Tompkins adventure by George Gamow (1939). In this paper we go through the different relativistic effects  (e.g. length contraction, time dilation, Doppler shift, light aberration) that show up during the boat trip and we explain how these effects were implemented in the $360^\circ$ video production process. We also provide exercise questions that can be used - in combination with the VR movie - to gain insight and sharpen the intuition on the basic concepts of special relativity. 
\end{abstract}

\maketitle
\section*{Introduction}
In the VR movie Captain Einstein \cite{Captain} you experience a boat trip in the beautiful city centre of Ghent in a world with a slow speed of light $c= 20km/h$. Starting at a small initial velocity 
$v$, different accelerations bring you finally up to lightspeed. During the various stages ($v/c\approx 40, 70, 85, 95\%,...$) you can see how the different effects of special relativity emerge gradually. An obvious advantage with respect to the real world is that these effects now become more tangible as they are brought to a human scale. This was the original idea of Gamow that led to the intriguing short story \emph{Mr. Tompkins} (see figure 1), which is set in such a dreamworld with a slow speed of light \cite{Gamow}. Later approaches in the same spirit include the nice animations of a relativistic bicycle trip through the city of T\"{u}bingen \cite{Tubingen,Tubingen2}; \emph{Real Time relativity} \cite{RTR, Realtime, Realtime2}, an interesting first-person game that allows you to travel near the speed of light in different sci-fi settings; and \emph{A Slower Speed of Light} \cite{mit,mit2,mit3}, an engaging first-person relativity game which takes place in a fantasy world. See also \cite{Weiskopf} for a thorough study on the visualization of both special and general relativity.

The new aspect of our project lies in the $360^\circ$ technology. First of all it allowed us to record and use real images for creating our movie, as opposed to a 3D virtual model. This clearly helped for the realistic 'feel' (but see also section II for a discussion on the limitations of this approach). In addition, the $360^\circ$ VR viewing experience puts the user 'inside the theory', so to speak. Not only does this lead to a powerful immersive experience, but it also allows for a very natural way to examine the directional dependence of the relativistic effects, notably of the Doppler shift. See also \cite{dome} for a recent spin-off of \cite{mit,mit2,mit3}, involving a dome projection in a planetarium. In the broader context of general relativity we should also mention the recent VR movies of black hole adventures \cite{VRBH1,VRBH2}.

Captain Einstein has so far been mainly used for science communication purposes. It has featured at festivals in Belgium and the Netherlands, drawing much interest and provoking enthusiastic reactions from people of different ages and backgrounds. But the origin of the movie is actually more educational: it grew out of the practice of teaching the theory of relativity at Ghent University. The main purpose of this paper is to provide the scientific content behind the movie, which can be used directly by students, or by lecturers in the context of an exercise session on special relativity.\footnote{We refer to our website \cite{Captain} for a less technical discussion.}  To this end much of the discussion is put in the form of exercise questions. In fact, as will become clear in this paper, one can think of the whole Captain Einstein project as one big exercise in special relativity. At the end of the paper, we also briefly comment on our experience with the movie so far, both in the context of science popularisation and education.  

Although it clearly would have been easier for Einstein to discover the theory of relativity in a world with a slow speed of light, deducing the laws of relativity from the direct observations still requires some work. Even to somebody with a prior training in relativity it is not entirely straightforward to relate our VR experience directly to the archetypical relativistic effects. The reason lies in the important difference between {\em what is} and {\em what is seen}: since the speed of light is finite we do not see fixed-time snapshots, the further the object the more we see it in the past. In more technical terms: at each instant we see our past light cone rather than a particular spacelike hypersurface. This might seem to complicate the use of the movie for illustrating the basic effects of relativity, but as we hope to convey in this paper, thinking about the true observational consequences of e.g. length contraction and time dilation can actually help to sharpen the understanding of these effects. We refer the reader to \cite{Lampa, Penrose, Terell} for the original groundbreaking work on the visual consequences of relativity. 

The science covered by our VR movie is basically the content of Einstein's 1905 paper\cite{Einstein} (apart from the Lorentz transformations of Maxwell's equations). In section I we study the role of length contraction, time dilation and relativistic velocity addition in our movie. In section II we examine the actual observations in a world with a slow speed of light. This involves taking into account the relativistic light aberration. As we will show, the light aberration formula also goes to the heart of our $360^\circ$ simulation. Finally, in section III we consider the relativistic Doppler shift. To our knowledge, this is the first time that the Doppler shift on realistic full light spectra of the sky has been simulated in this detail. This simulation produced the spectacular colour effects in the movie with for instance the rainbowlike features that can be traced back to the different greenhouse absorption bands in the IR. 

The exercises in the different sections are aimed at the level of someone with a rudimentary knowledge of special relativity, e.g. someone who has attended the first few lectures of a (special) relativity class or has read the first few chapters of some special relativity course (see for instance \cite{hogg} for some excellent online notes). The stars (one or two) indicate the difficulty level, exercise solutions are provided at the very end of the paper. 

To set the conventions: we will consistently use primes (e.g. $t',x', \theta',\lambda',\ldots $) for coordinates, directions, wavelengths,... in the boat reference frame $S'$. This is your reference frame when you experience the VR boat trip. Non-primed coordinates, directions,... are reserved for the quay rest frame $S$.

\section{Mr. Tompkins}    
\begin{figure}
\begin{center}
\includegraphics[width=.47\textwidth]{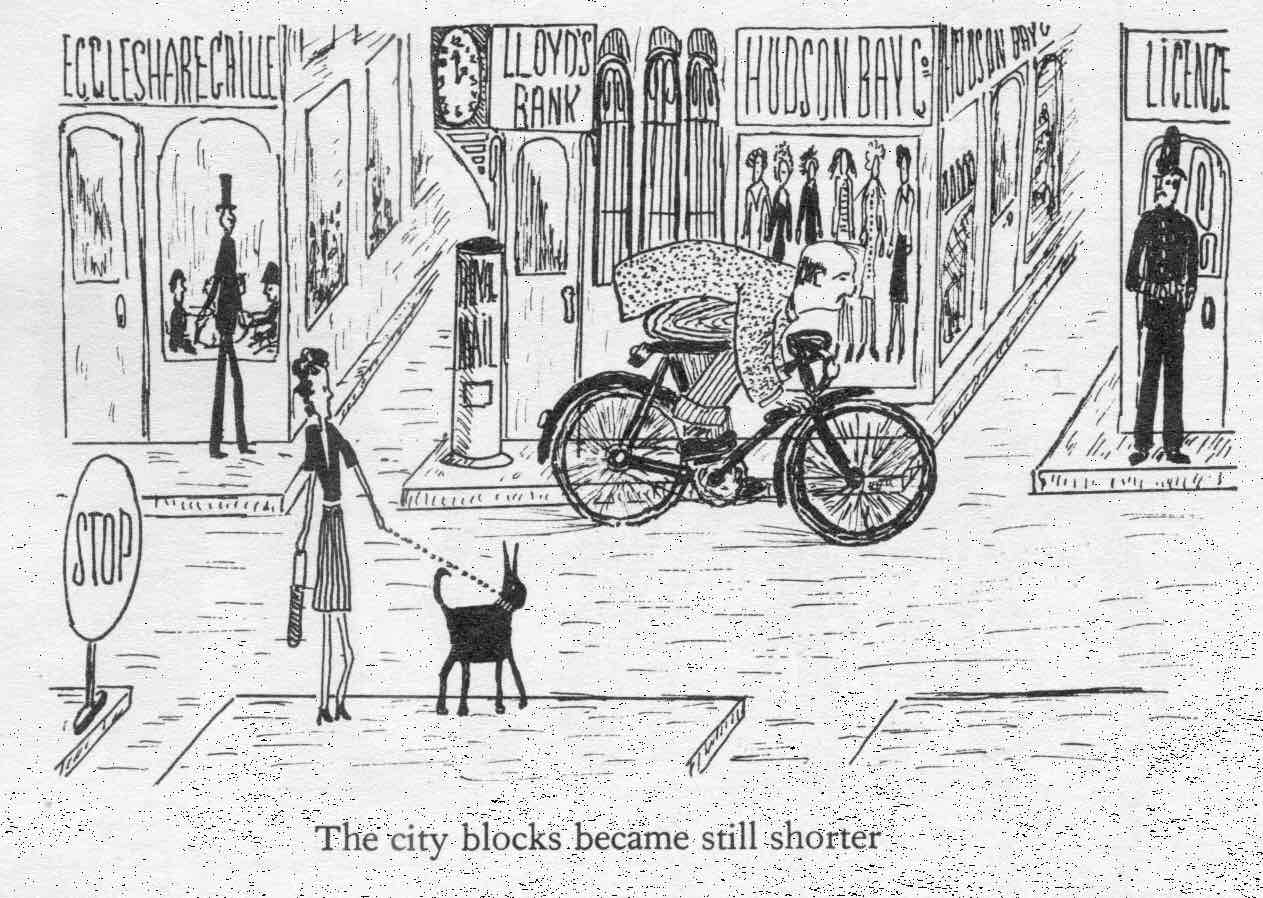}
\captionsetup{justification=raggedright}
\caption{Mr. Tompkins (Gamow, 1939) racing through the streets in a length contracted world.}
\end{center}
\label{tompkins}
\end{figure} 

In 1939 the brilliant physicist George Gamow writes the delightful science popularisation book {\em Mr. Tompkins in wonderland} \cite{Gamow}, in which Mr. Tompkins wakes up in a dreamworld with a low speed of light ($c=10 mph$), much like the one we have in Captain Einstein. In this world Mr. Tompkins is confronted with the effects of length contraction and time dilation. In figure 1 for instance,  you see Mr. Tompkins racing through the streets, with the buildings (and people on the sidewalk) experiencing a length contraction.

\begin{figure}[t]
\begin{subfigure}[b]{.45\textwidth}
\includegraphics[width=\textwidth]{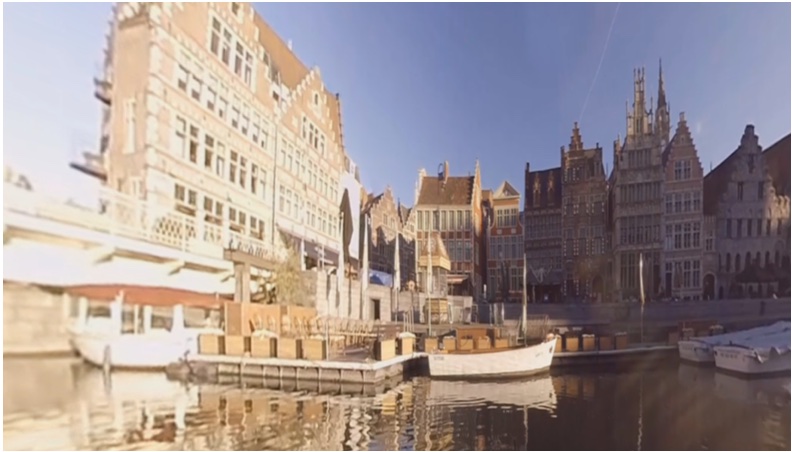}
\vspace{0.1cm}
\end{subfigure}\hfill
\begin{subfigure}[b]{.45\textwidth}
\includegraphics[width=\textwidth]{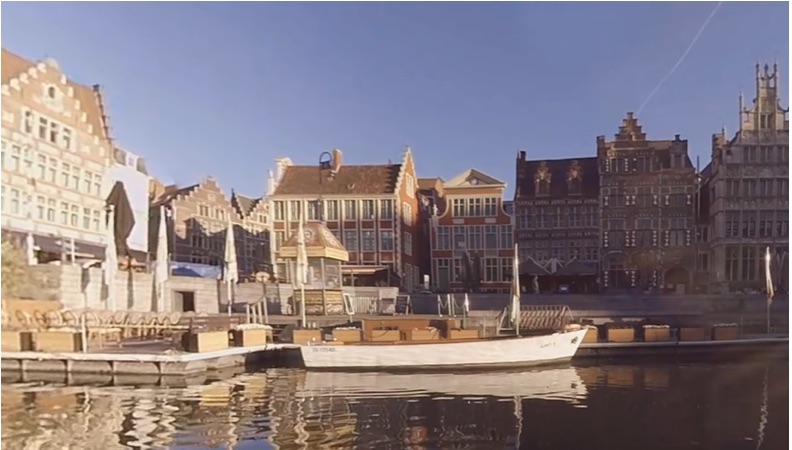}
\end{subfigure}\vskip\baselineskip
\captionsetup{justification=raggedright}
\caption{Above: point of view shot from the boat orthogonal to the velocity direction for a boat trip along a length contracted Graslei at speed $v=0.85c$. Below: image in the quay rest frame ($v=0$) for comparison.}
\label{figlengthcontraction}
\end{figure}

The time dilation and length contraction obviously also play an important role during Captain Einstein's boat trip. In figure {\ref{figlengthcontraction}} you see the analogous picture for a boat trip along the canals of a length contracted Ghent at a velocity $v= 0.85 c$. From the boat's perspective it is of course the quay that is moving, which results in a length contraction $l'=l/\gamma(v)$ along the direction of movement for all the objects at rest in the quay frame. Here we define the Lorentzfactor $\gamma(v)\equiv 1/\sqrt{1-v^2/c^2}$ and for $v=0.85 c$ we have $\gamma(v)\approx 2$, corresponding to the situation in the figure. From the Captain Einstein movie you already know that this is not the actual view, but to be sure, the picture \emph{does} represent the true geometry of the moving quay from the perspective of the boat observer $O'$ at some fixed time $t'$. We reserve the discussion of the visual manifestation of length contraction for the next section, while the visual consequences of time dilation are discussed in section III.  In this section you are asked to compute the effect of length contraction and time dilation on the boat travel time, which in turn has its consequences for our simulation of speed.  \vspace{0.2cm}

\noindent {\bf Exercise 1 ($\star$): Length contraction, time dilation and frame rates.}

(a) Let us take two subsequent bridges over the canal that are separated by a distance $l$ in the quay rest frame. Given the length contraction, if the boat is going at constant speed $v$, what is the time $\Delta t'$ it takes the boat to drive from one bridge to the next?\vspace{0.1cm}

(b) From the perspective of the quay observer there is of course no length contraction for the distance between the two bridges. Still he should be able to compute the boat travel time $\Delta t'$ and arrive at the same conclusion. Show that, by taking into account the time dilation, the quay observer indeed arrives at the same result for $\Delta t'$. This should clarify Maja's remark: "\emph{we get some speed for free, time is slowing down as we speed up.}"  \vspace{0.1cm}

(c) The original footage was shot on a boat going at constant velocity $v_b(\approx 8km/h)$. To simulate other velocities we simply applied a speed-up to the movie.\footnote{See for instance this \href{https://youtu.be/v8dWW9KxqlQ?t=47s}{\underline{clip}} for the same speed-up trick in the legendary 80s TV series Knight Rider.} In a non-relativistic world we would have to speed up the movie by a factor $v/v_{b}$ to simulate a certain velocity $v$. What is the speed-up that we had to apply for our relativistic movie?$\rfloor$\vspace{0.2cm}

\noindent{\bf Exercise 2 ($\star \star$): "\emph{This g-force is crushing us,}" relativistic velocity addition and acceleration}

The different accelerations during the boat trip bring you from one instantaneous inertial frame to the next: $S'=S(\vec{v})$, where $S(\vec{v})$ is the inertial frame that moves with a velocity $\vec{v}$ with respect to the quay frame $S=S(\vec{0})$. The movie shows you the visual effects of these accelerations, (un)fortunately we can not let you experience the corresponding g-forces. But what is the g-force $g$, corresponding to an acceleration $a=\frac{d v}{d t'}$? (Here $t'$ is the time-coordinate for the boat observer $O'$! So $a$ is the acceleration that we measure on the boat by timing the change of speed on our speedometer.) For simplicity you can assume a constant direction: $\vec{v}(t')=(v(t'),0,0)$.  

At the end of the Captain Einstein movie the boat is accelerating out of control towards lightspeed. We hear Maja shouting: "\emph{we have reached 19.8\ldots19.9 km/h, this g-force is crushing us.}" Assuming a constant acceleration $a$ of about 0.1km/h in one second, $a\approx 0.003 \,G$ and a maximal physiologically sustainable g-force of 10G, at what velocity would you pass out?$\rfloor$

\begin{figure}
\includegraphics[width=0.46\textwidth]{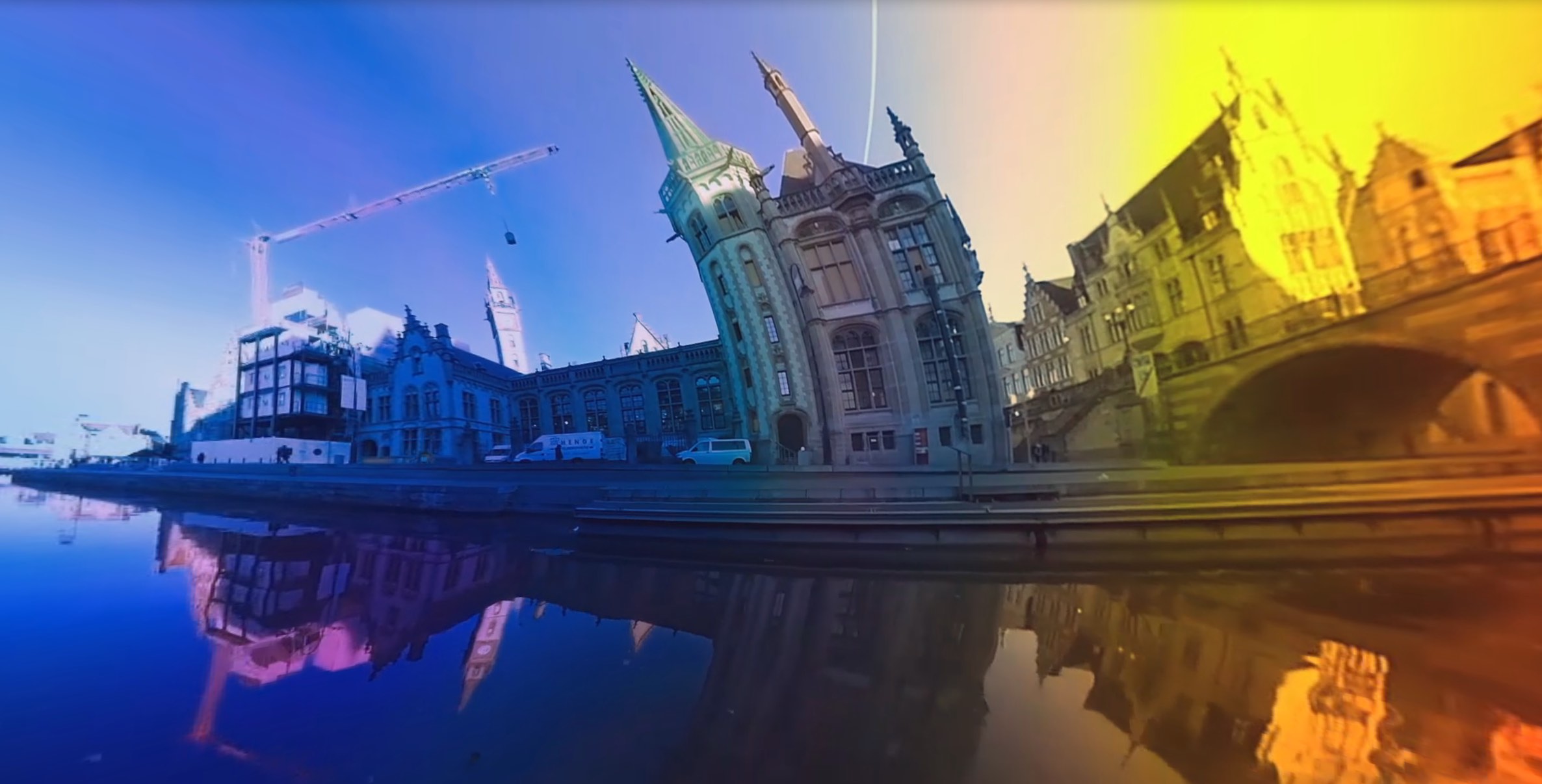}
\captionsetup{justification=raggedright}
\caption{"\emph{Check out the towers, how they are curved!}" Point of view shot from starboard side (righthand side) in the Captain Einstein movie.}
\label{figtowers}
\end{figure}

\section{Observing on the light cone}
The actual shapes of the buildings in the Captain Einstein movie (see e.g. figure \ref{figtowers}) are quite different from what you see in figure \ref{figlengthcontraction}. Length contraction refers to the length (along the velocity direction) of moving objects taken at a fixed time $t'$ in the considered reference frame. However, since the speed of light is finite, we do not see fixed-time snapshots. At each instant we see our past light cone, rather than a particular spacelike hypersurface. This \emph{vision delay effect} was neglected in the Mr. Tompkins book. \vspace{0.2cm}

\noindent{\bf Exercise 3 ($\star$): "\emph{Check out the towers!}"} 

For the tower in figure \ref{figtowers}, it is clear that the light travelling from the top of the tower towards our eye, has taken a longer time than the light that emerged from the base of the tower. Use this to explain qualitatively the distortion in the picture, in particular the direction in which the tower bends.$\rfloor$ \vspace{0.2cm}

\noindent{\bf Exercise 4 ($\star$): How to see the length contraction}

At some point during the trip Maja asks you casually to "\emph{see how the ancient houses are squeezed to half their width}." She is of course referring to the length contraction, at $17km/h$ we have $\gamma(v)\approx 2$, which indeed gives a contraction by a factor of two. But she was a bit cheeky there: to really \emph{see} the length contraction you have to look in a very specific direction. 

\begin{figure}
\begin{subfigure}[b]{.45\textwidth}
\includegraphics[width=\textwidth]{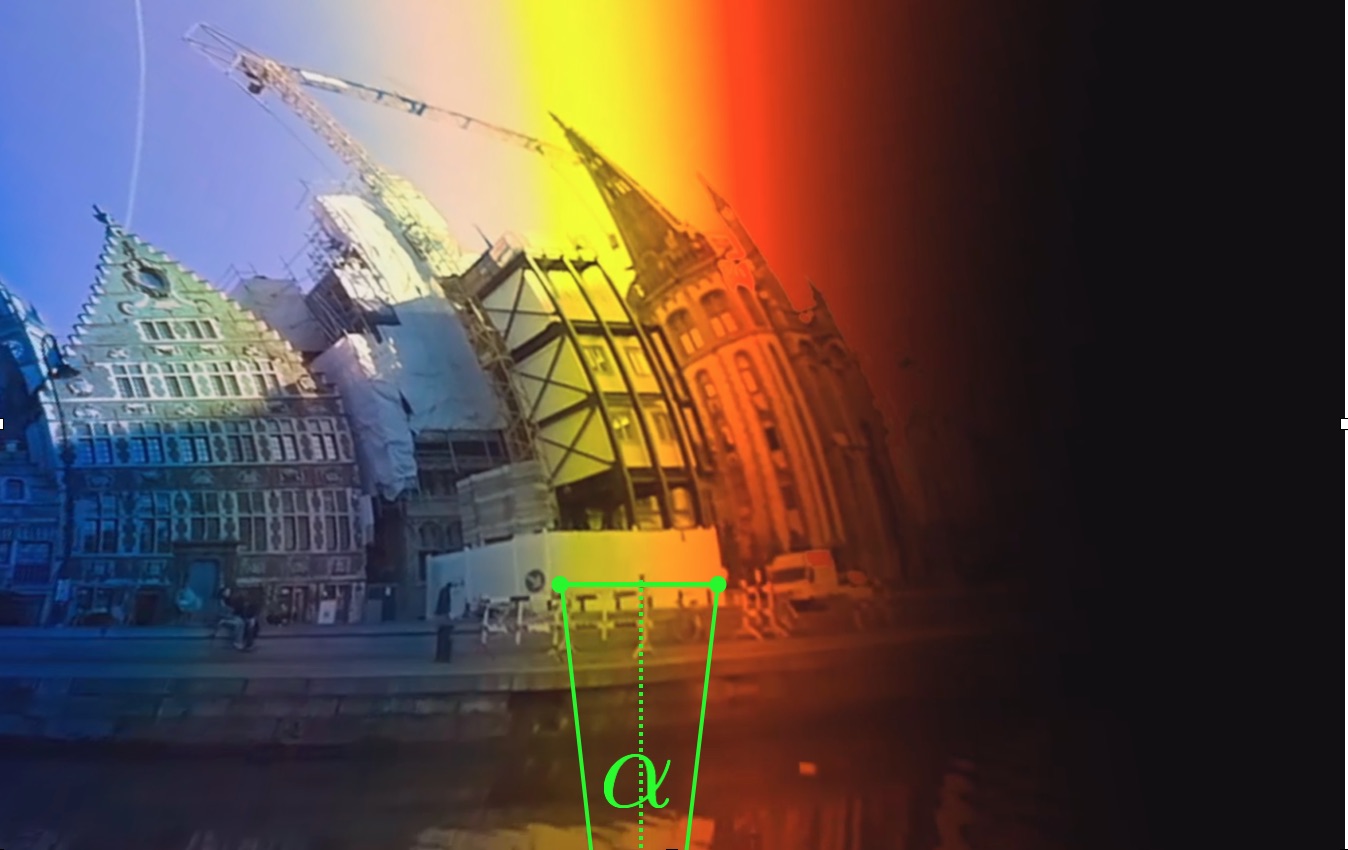}
\vspace{0.1cm}
\end{subfigure}\hfill
\begin{subfigure}[b]{.45\textwidth}
\includegraphics[width=\textwidth]{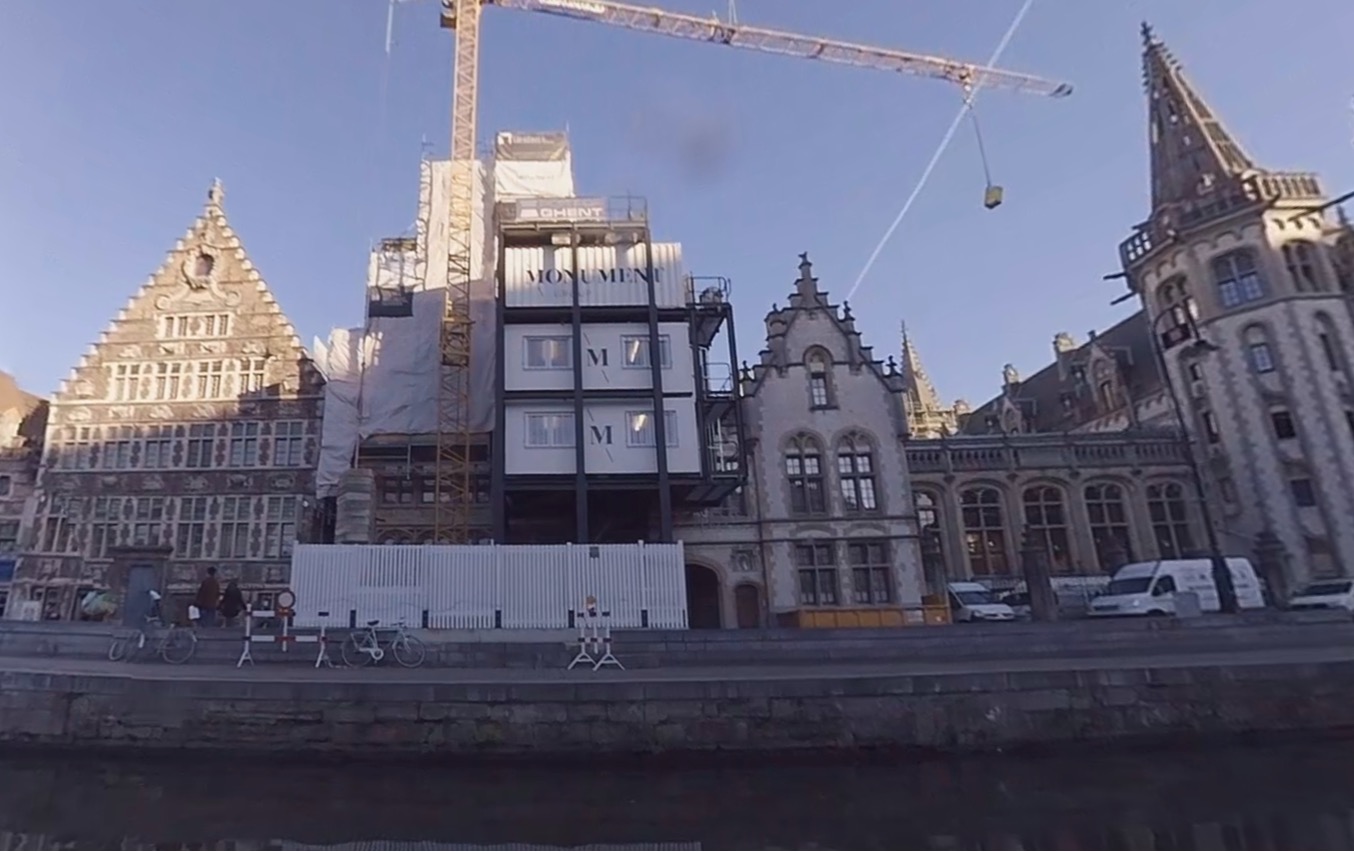}
\end{subfigure}\vskip\baselineskip
\captionsetup{justification=raggedright}
\caption{Above: length contracted white fence as seen at starboard by Captain Einstein (the fence lies parallel to the velocity direction). The dotted line is the $\theta'=\pi/2$ direction orthogonal to the velocity ($\vec{v}.\vec{n}'= v \cos \theta'$ with $\vec{n}'$ the unit vector in a particular viewing direction). Below: view in the quay rest frame for comparison.}
\label{figfence}
\end{figure}

Explain that by looking orthogonally to $\vec{v}$ one can indeed directly observe the length contraction, without distortions from the vision delay effect. So if we take for instance a picture in the orthogonal direction, precisely at the moment when the moving object (for instance the white fence in figure \ref{figfence}) appears in the centre of the viewfinder, we can use standard Euclidean geometry to compute its length $l'$ from the opening angle $\alpha$ and the distance between the boat and the object (see also the left panel of figure \ref{figabcon}).

Now also argue qualitatively from the vision delay effect why the house on the left in the picture does not appear to be contracted and why the white van on the right appears to be more contracted.$\rfloor$\vspace{0.2cm}

\begin{figure}[t]
\begin{subfigure}[b]{.45\textwidth}
\includegraphics[width=\textwidth]{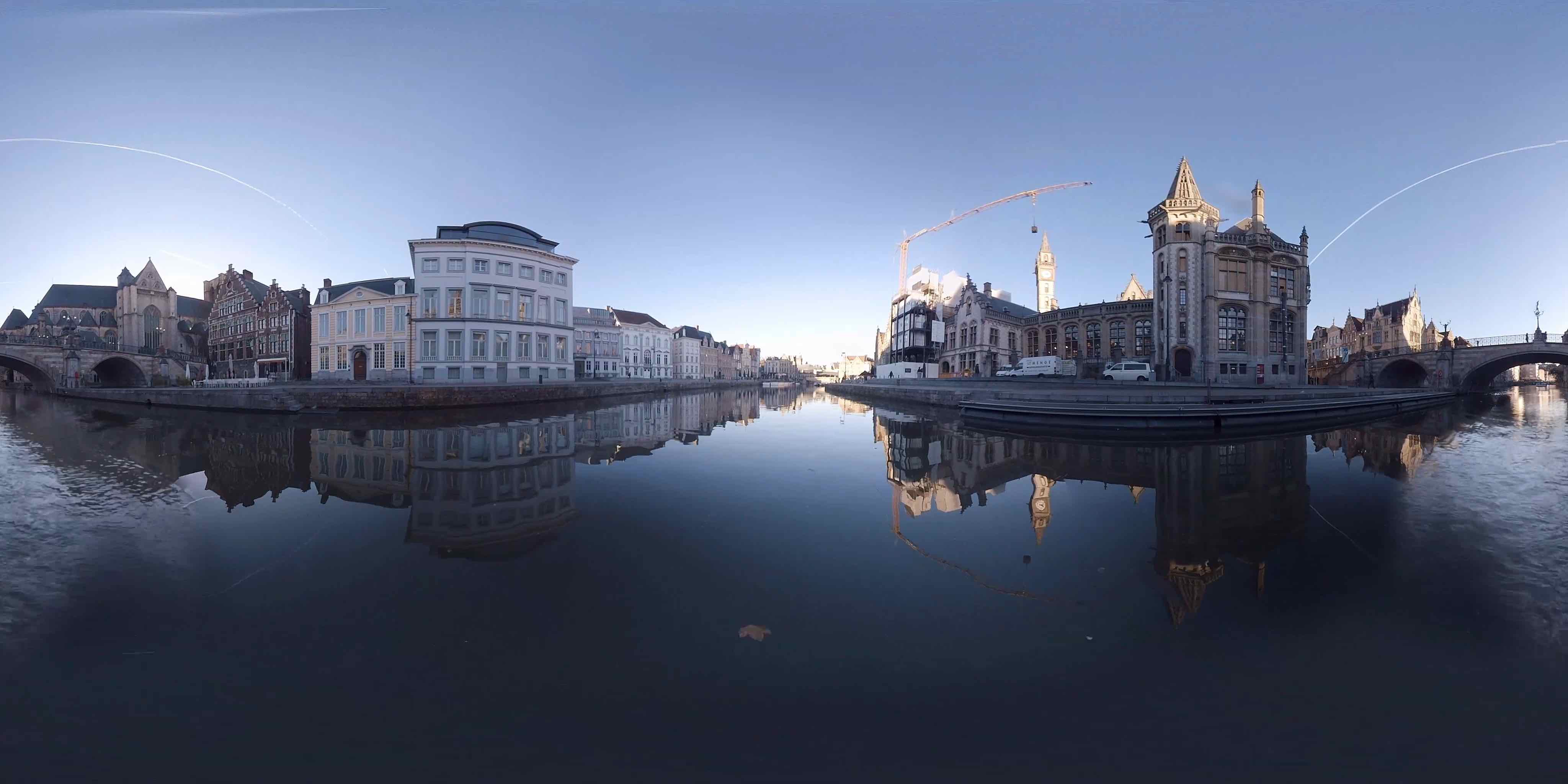}
\caption{}
\vspace{0.3cm}
\end{subfigure}\hfill
\begin{subfigure}[b]{.455\textwidth}
\includegraphics[width=\textwidth]{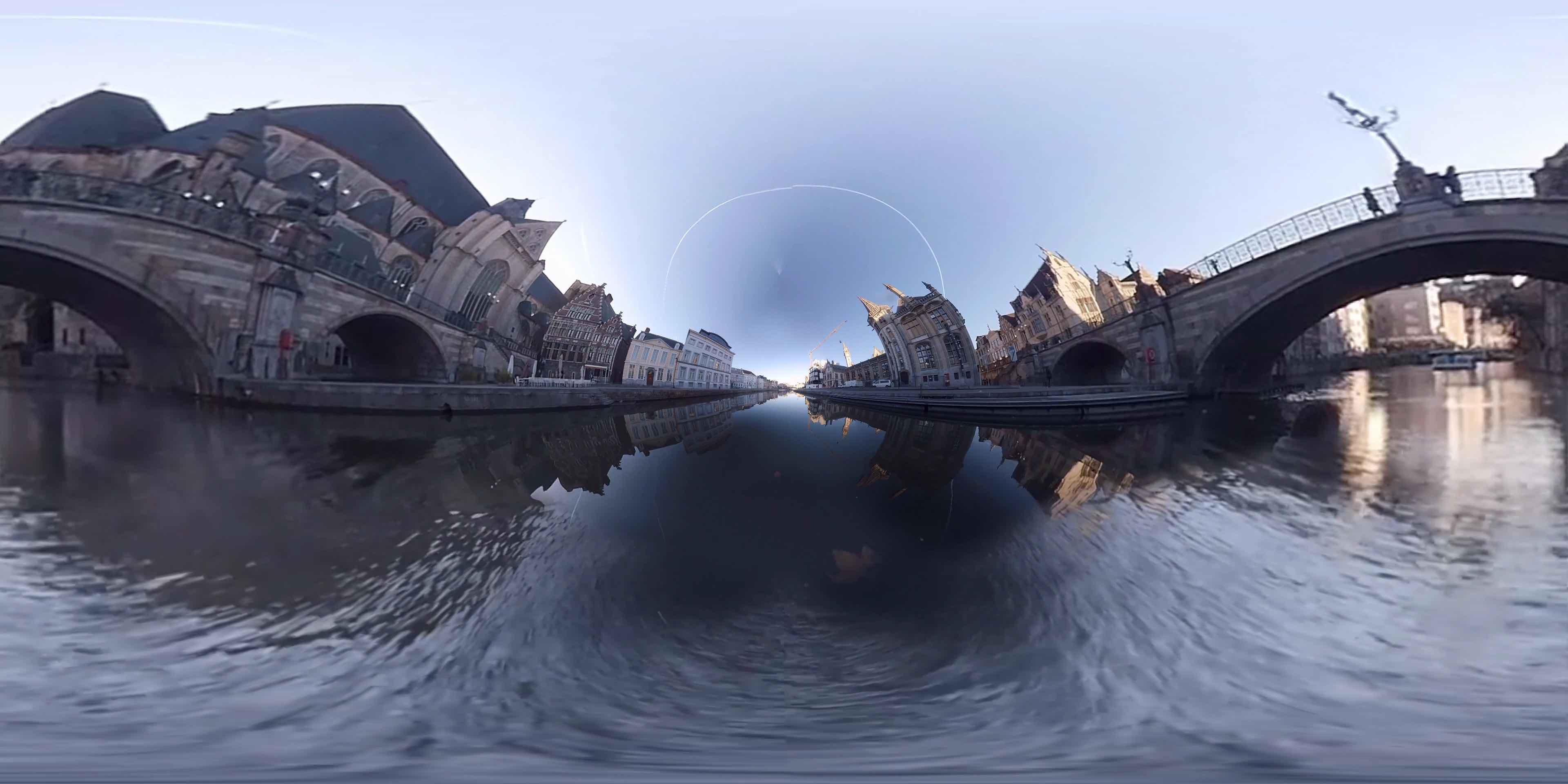}\caption{}
\end{subfigure}\vskip\baselineskip
\captionsetup{justification=raggedright}
\caption{Above (a): frame of the original footage of the full sphere of directions $\vec{n}$ in equirectangular projection: $\vec{n}=(\sin(\pi-v)\cos(u),\sin(\pi-v)\sin(u),\cos(\pi-v))\,,$ with $u\in]\pi,-\pi]$ the horizontal coordinate and $v\in[0,\pi]$ the vertical coordinate in the picture. The boat is going in the $x$-direction: $\vec{v}=(v,0,0)$. Below (b): corresponding frame after applying the light aberration formula (\ref{deform}) for $v=0.7c$. {\bf Side question}: can you understand what happened with the plane trail in the picture? }
\label{figequi}
\end{figure}

Due to this vision delay effect, it seems that to make a relativistic movie one needs the full 4D information, in particular it seems one needs the 3D position for all the relevant objects at the appropriate times. We did not have this. What we did have is the series of different images recorded by a $360^{\circ}$ camera on the boat, during the shoot along the canals of Ghent. In other words: at each instant we know the light that enters the camera from every direction. As can be seen in figure \ref{figequi} (a), the resulting image is typically stored in an equirectangular projection of the unit sphere.

\begin{figure}[t]
\begin{subfigure}[b]{.20\textwidth}
\includegraphics[width=\textwidth]{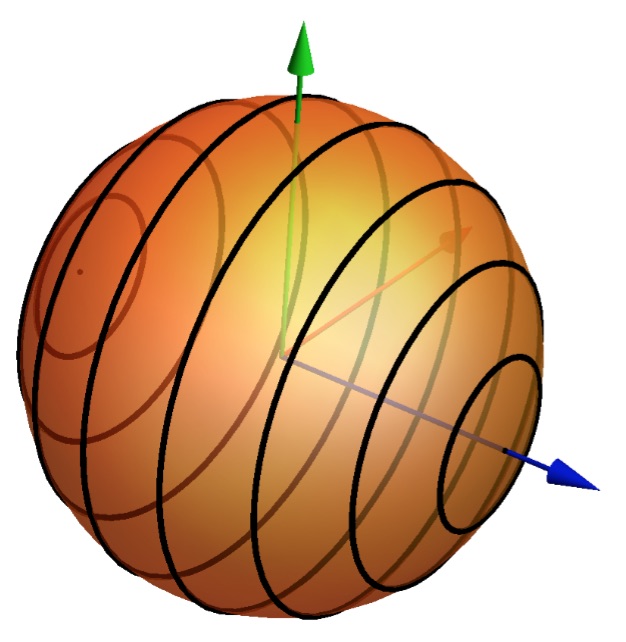}
\end{subfigure}\hfill
\begin{subfigure}[b]{.20\textwidth}
\includegraphics[width=\textwidth]{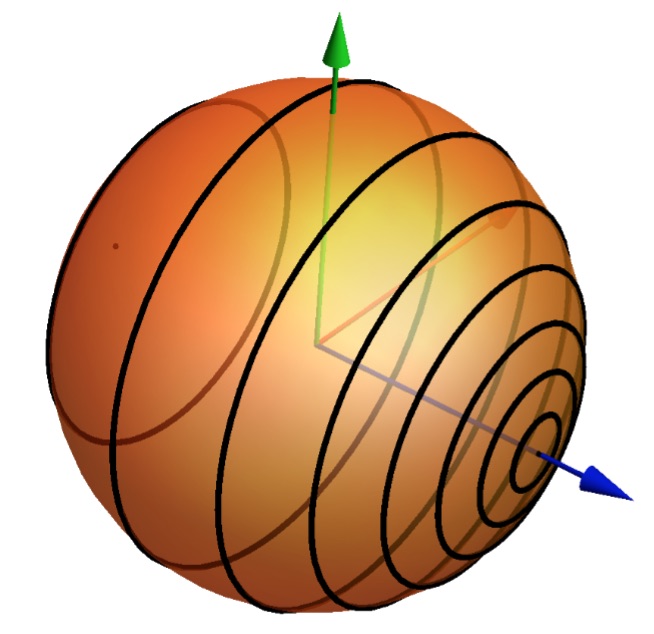}
\end{subfigure}\vskip\baselineskip
\captionsetup{justification=raggedright}
\caption{Visualization of the light aberration formula (\ref{deform}). Left: the unit-sphere of directions $\vec{n}$ in the quay rest frame. The different black circles are parallels with respect to the (blue) $x$-axis that each correspond to a particular angle $\theta$: $\vec{n}.\vec{e}_x=\cos\theta$. Right: the corresponding unit-sphere of directions $\vec{n}'$ ($\vec{n}'.\vec{e}_x=\cos\theta$), after a boost in the $x$-direction with $v/c=0.7$. As can be seen from the transformed parallels, the field of view 'collapses' into the velocity-direction.}
\label{aberration}
\end{figure}

Since our boat was going at $v_b\approx8km/h$ in the real world with $c\approx3\times 10^8m/s$, and thus $v_b/c\approx 0$, we can effectively consider each recorded movie frame to be the image in the quay rest frame $S$ of our virtual world with a slow speed of light. Notice, that for this to hold all the objects in the image have to be at rest in the quay frame. This is the reason why we had to stitch away our boat in the original images. This is also the reason why we  shot the movie on a cold winter day, we did not want any other moving boats in the image. Unfortunately the cold did not stop the traffic at the quay. So the people on foot, the bikes and the tram in the movie are not depicted correctly and can be considered as goofs that defy the laws of relativity (see exercise 8 for the correct visualisation of the pedestrians).

It is now important to realize that the very same photons that generate the $360^\circ$ image in the quay frame $S$, are also responsible for the image in any other frame $S'$. To create this new image we essentially have to know the directions and wavelengths of these photons, as observed in $S'$. Here we consider the directions, the wavelengths (Doppler shift) are discussed in the next section. The transformation of the unit sphere of light directions $\vec{n}$ in the quay rest frame $S$ to the unit sphere of directions $\vec{n}'$ in the observer boat frame $S'$ is given by the relativistic aberration formula, we provide it here for reference (with $\vec{v}.\vec{n}= v \cos \theta,\,\ \vec{v}.\vec{n}'= v\cos \theta'$): \be \cos\theta'=\frac{\cos\theta +\frac{v}{c}}{1+\frac{v}{c}\cos\theta}\label{deform}.\ee 
In figure \ref{aberration} we show a visualisation of the formula on the unit sphere, the resulting transformation on the $360^{\circ}$ footage is shown in figure \ref{figequi} (b). 

This transformation can be implemented fairly straightforward with a \emph{shader}, an image transformation program common in computer graphics software (e.g. WebGL, Unity), written in a specific GPU compatible language. We have experimented with that, but in the end, for our final movie we wrote a Matlab code that generates the relativistically transformed $360^{\circ}$ images, frame by frame. The main reason lies in our implementation of the Doppler shift (see the next section), this required numerical interpolation on a large set of pre-generated RGB to RGB transformations, which we found more convenient to implement with Matlab.           

\vspace{0.1cm}

\noindent {\bf Exercise 5 ($\star\star$): from 2D to 3D, from light aberration to length contraction} 

So to create our movie we used a local transformation (\ref{deform}) for the 2D unit-sphere of light directions hitting the observer at a certain time. One can verify in general that the light aberration formula indeed reflects two effects in the 4D world: the vision delay effect and the length contraction.
In this exercise we focus on the particular situation of the previous exercise 4 and demonstrate that for an orthogonal viewing direction $\theta'=\pi/2$, the aberration formula is indeed geometrically equivalent to the length contraction. As shown in exercise 4, for this particular viewing angle, we can use Euclidean geometry to relate the opening angle $\alpha$ of the two endpoints to a length $l'$ of the moving fence: $l'=2 y \tan(\alpha/2),$ where we take the object lying parallel to $\vec{v}$ and $y$ is the distance from the observer to the centre of the object, see figure \ref{figabcon}. Now use the aberration formula to show that the length $l$ in the object's rest frame (i.e. the quay rest frame $S$) is indeed longer by a factor $\gamma(v)$.$\rfloor$

\begin{figure}[h]
\includegraphics[width=0.5\textwidth]{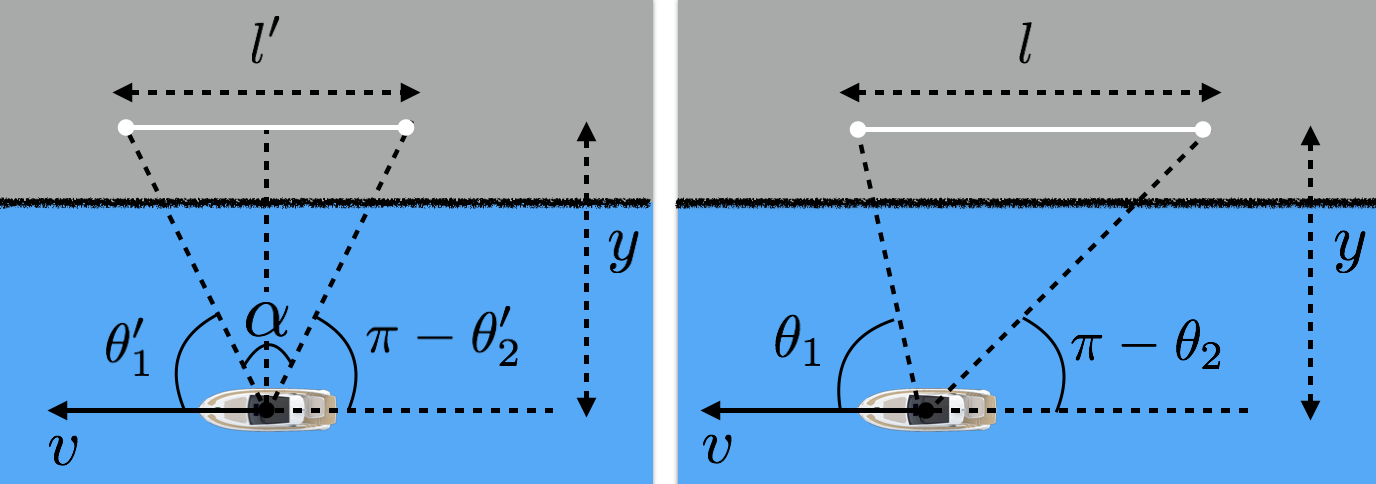}
\captionsetup{justification=raggedright}
\caption{Left: fence as seen by the boat observer, with the endpoints at angles $\theta'_1=\pi/2-\alpha/2$, $\theta'_2=\pi/2+\alpha/2$ with respect to the velocity vector. Right: the same fence as it would be seen in the quay rest frame (also the fence rest frame) for an observer at the same position as the boat observer. }
\label{figabcon}
\end{figure} 
\vspace{0.1cm}

\begin{figure}[t]
\begin{subfigure}[b]{.45\textwidth}
\includegraphics[width=\textwidth]{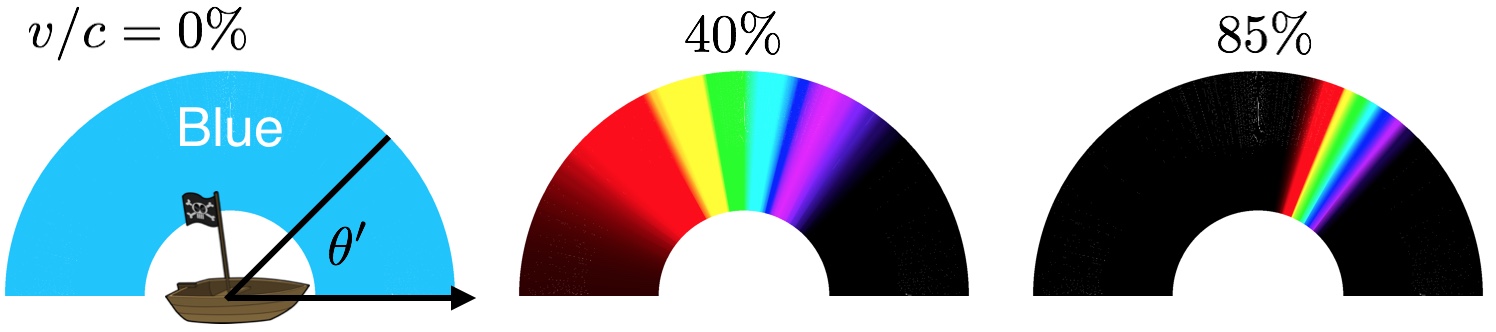}
\captionsetup{justification=raggedright}
\caption{ }
\vspace{0.3cm}
\end{subfigure}\hfill
\begin{subfigure}[b]{.45\textwidth}
\includegraphics[width=\textwidth]{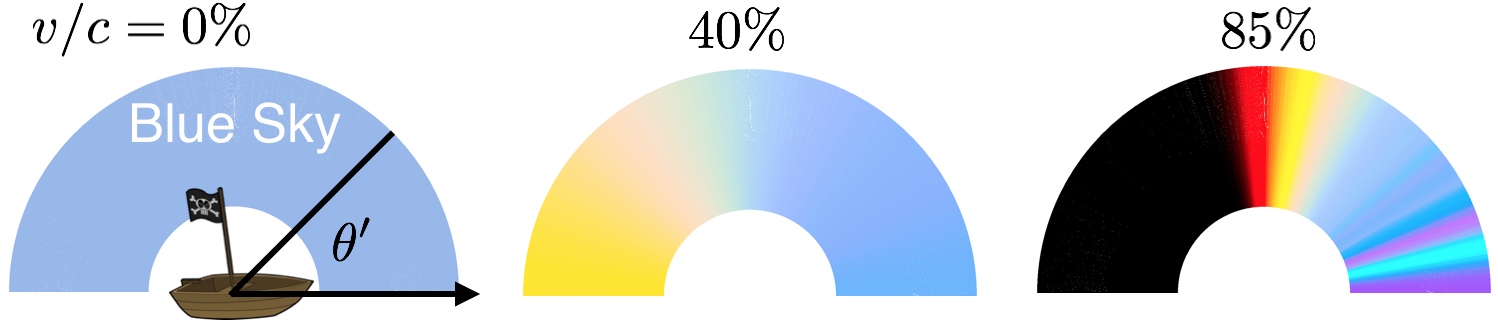}\caption{}
\end{subfigure}\vskip\baselineskip
\captionsetup{justification=raggedright}
\caption{(a) Doppler shifted monochromatic blue, $\lambda=475 nm$, at 40\% and 85\% of lightspeed. (b) Doppler shift for the diffuse blue sky spectrum that was used as input for the Captain Einstein movie (see figure \ref{figDoppmovie} for a point of view shot).}
\label{figdopp}
\end{figure} 

\section{Doppler extravaganza}
Let us now discuss colour. The relativistic Doppler formula reads (with again $\theta'$ the direction in  $S'$ with respect to the velocity direction): 
\be \lambda'=\lambda \, \frac{1-\frac{v}{c}\cos\theta'}{\sqrt{1-\frac{v^2}{c^2}}}\equiv \frac{\lambda}{D}\,. \label{doppform}\ee
We first illustrate the Doppler formula by looking at its effect on monochromatic light. In figure \ref{figdopp}(a) you see the effect on blue light of $\lambda=475 nm$, for $v/c=40\% \,(85\%)$. At these relativistic speeds, the Doppler shifted wavelengths span the full visible spectrum $\approx [390nm, 700nm]$ \cite{photopic}, and even go beyond the visible spectrum, into the UV: $\lambda'<390nm$  for directions $\theta'<52^{\circ}$ ( $\theta'<48^{\circ}$) and into the IR: $\lambda'>700nm$  for directions $\theta'>151^{\circ}$ ( $\theta>75^{\circ}$); resulting in black patches of invisible light both behind and in front of the boat. 

\begin{figure}
\includegraphics[width=0.48\textwidth]{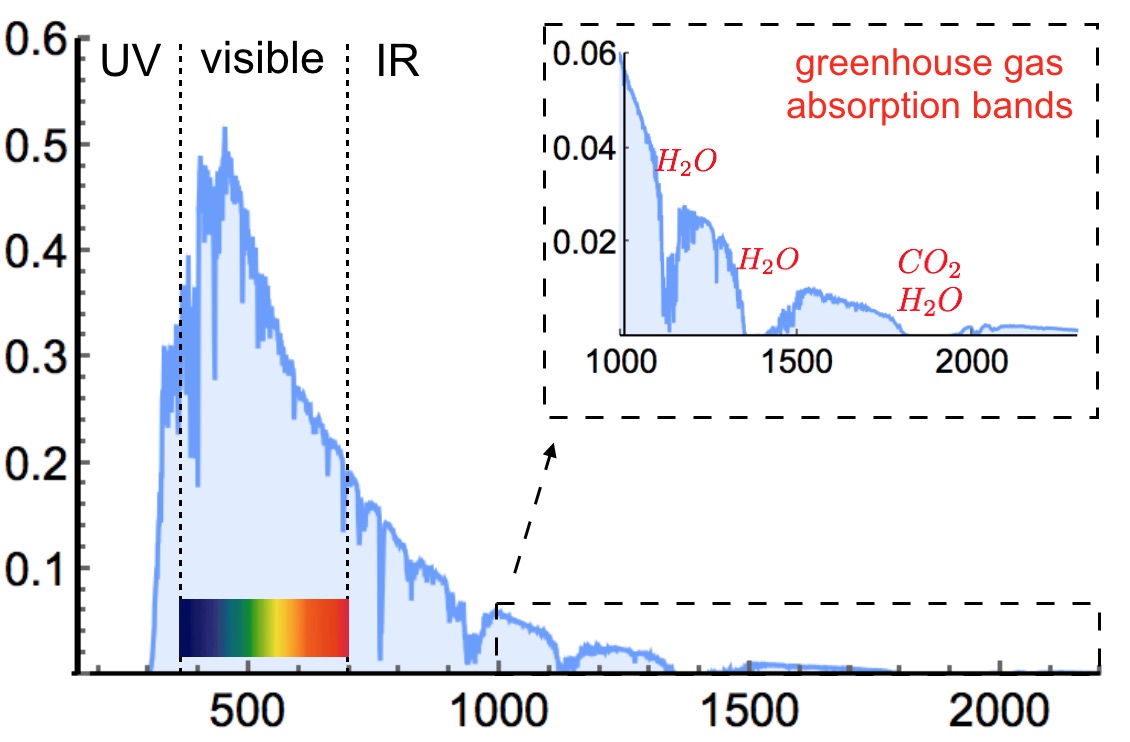}
\captionsetup{justification=raggedright}
\caption{The diffuse sky radiation (data from \cite{Smarts}). Horizontal axis: wavelengths in units $nm$; vertical axis: spectral power distribution $I(\lambda)$ in units $W/(m^2 \,nm)$. Inset: part of the IR tail with the greenhouse gas absorption bands.}
\label{figspectrum}
\end{figure}

\begin{figure}[t]
\begin{center}
\includegraphics[width=0.48\textwidth]{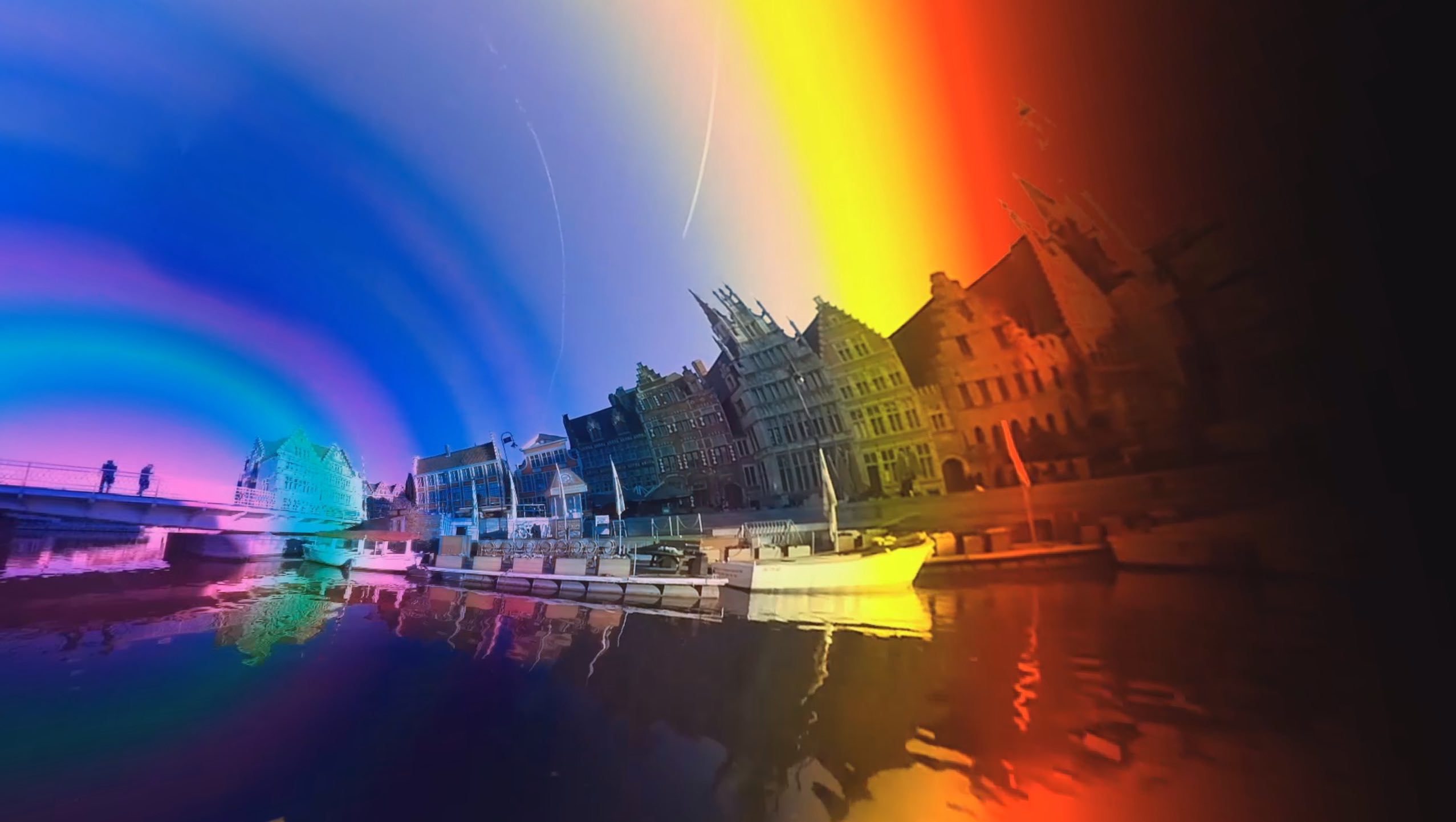}
\captionsetup{justification=raggedright}
\caption{Point of view shot from Captain Einstein at $v=0.85 c$, taken at an angle $\theta'\approx 30^{\circ}$ with respect to the velocity direction.}
\label{figDoppmovie}
\end{center}
\end{figure}

The considered values of $v/c$ for the Doppler shift on blue light in figure \ref{figdopp}(a)  correspond to the velocity at the beginning of our movie  $(8km/h=0.4 c)$ and to the velocity after the second acceleration ($17km/h=0.85c$).  Yet the colours at the sky in our movie do not look at all like in the figure above! The reason lies in the difference between the pure blue light and the incredible rich blue sky that consists of a full spectrum of different wavelengths, starting in the UV and ending deep in the IR part of the light spectrum (see figure \ref{figspectrum}). The Doppler effect on such a realistic blue sky spectrum produces the colour pallet of figure \ref{figdopp}(b), see also figure \ref{figDoppmovie} for a point of view shot in our movie. At the end of this section we will expand a bit more on our simulations of the Doppler shift, but now you are first asked to explain the qualitative features of the colours in the movie and think about the manifestations of time dilation. \vspace{0.2cm}   

 \vspace{0.7cm}

\noindent {\bf Exercise 6 ($\star$): Captain Einstein and the Doppler shift: seeing the invisible} 

In this exercise we focus on the $v=0.85c$ case (figure \ref{figdopp}(b), right panel). At this relativistic speed the Doppler shift allows you to see parts of the blue sky spectrum (figure \ref{figspectrum}) that normally lie beyond our visible window. Can you explain the yellow and red bands at orthogonal viewing directions $\theta'\approx 90^{\circ}$? In the movie Maja reassures us: "\emph{don't worry about the dark behind you, it's not a black hole.}" Can you explain the true reason for this darkness? Finally, can you see why the greenhouse gas absorption bands are responsible for the distinct blue and purple bands in front of the boat? (Don't worry here about the overall brightness in front of the boat, as we explain below, the searchlight effect boosts the light intensity in the frontal directions.)  $\rfloor$

\vspace{0.2cm}

\noindent {\bf Exercise 7 ($\star$): How to see the time dilation: transverse Doppler effect} 

The relativity of time, and in particular the time dilation, is probably the most spectacular consequence of the theory of relativity. In section I we discussed the effect of time dilation on the experience and simulation of speed in our movie. Now we discuss other manifestations. The most direct verification of time dilation results from comparing two clocks after different trips \cite{hafelekeating}. Our captain Maja Einstein is alluding to this type of twin paradox experiment. In our movie she is thirtysomething, while her brother died in 1955. But this is of course a bit of a gimmick. However, you can also directly \emph{see} the time dilation during the boat trip. For this you simply have to look at the sky. Indeed, the relativistic Doppler effect is in fact the combination of two phenomena: the (de)compression of the light waves due to the radial motion of the source with respect to the observer (classical Doppler effect, numerator of eq. (\ref{doppform})) + the time dilation and therefore frequency shift due to the relative velocity of the source (transverse Doppler effect, denominator of eq. (\ref{doppform})). At angles $\theta'=\pi/2$ we see that part of the sky for which the radial velocity is zero. The Doppler shift in those directions is therefore purely the result of time dilation: the 'sky clock' that is ticking at a slower rate, resulting in an observed light frequency that is smaller by a factor $\sqrt{1-v^2/c^2}$.  For this exercise we simply ask you to put on the VR headset and verify this transverse Doppler effect during the Captain Einstein boat trip.$\rfloor$            
 \vspace{0.2cm}
 
\noindent {\bf Exercise 8 ($\star $): Dopplerganger: from fast to slow motion}

\begin{figure}[t]
\begin{center}
\includegraphics[width=0.39\textwidth]{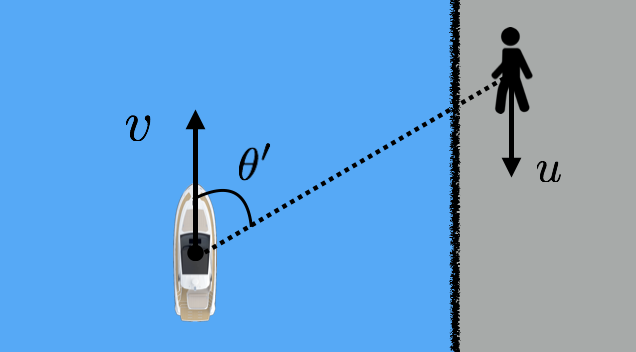}
\captionsetup{justification=raggedright}
\caption{Pedestrian walking with velocity $\vec{u}=-u \vec{e}_x$ in the quay frame, viewed at an angle $\theta'$ in the boat frame. The boat has velocity $\vec{v}=v \vec{e}_x$ in the quay frame.}
\label{figdopplerganger}
\end{center}
\end{figure}

During our boat trip you can see several pedestrians that are speed walking in 'silent movie style'. As we explained in section II, our procedure can not depict correctly the objects (like pedestrians) that are in motion with respect to the quay. In this exercise we examine what the observed walking pace should have been. We consider somebody walking along the quay with a velocity $\vec{u}$ (see figure \ref{figdopplerganger}) in the quay frame. 

(a) What is the velocity $u'=-dx'/dt'$ of our pedestrian friend in the boat frame?

(b) In her eigenframe the pedestrian is walking at a pace $f_0$ of one step per second. What is then the walking pace $f'$ that we would observe on the boat as a function of our viewing direction $\theta'$?  For what angles $\theta'$ would we observe the pedestrian in slow (fast) motion: $f'<f_0\,\, (f'>f_0)$.$\rfloor$
\vspace{0.2cm}


\begin{figure}[b]
\begin{center}
\includegraphics[width=0.45\textwidth]{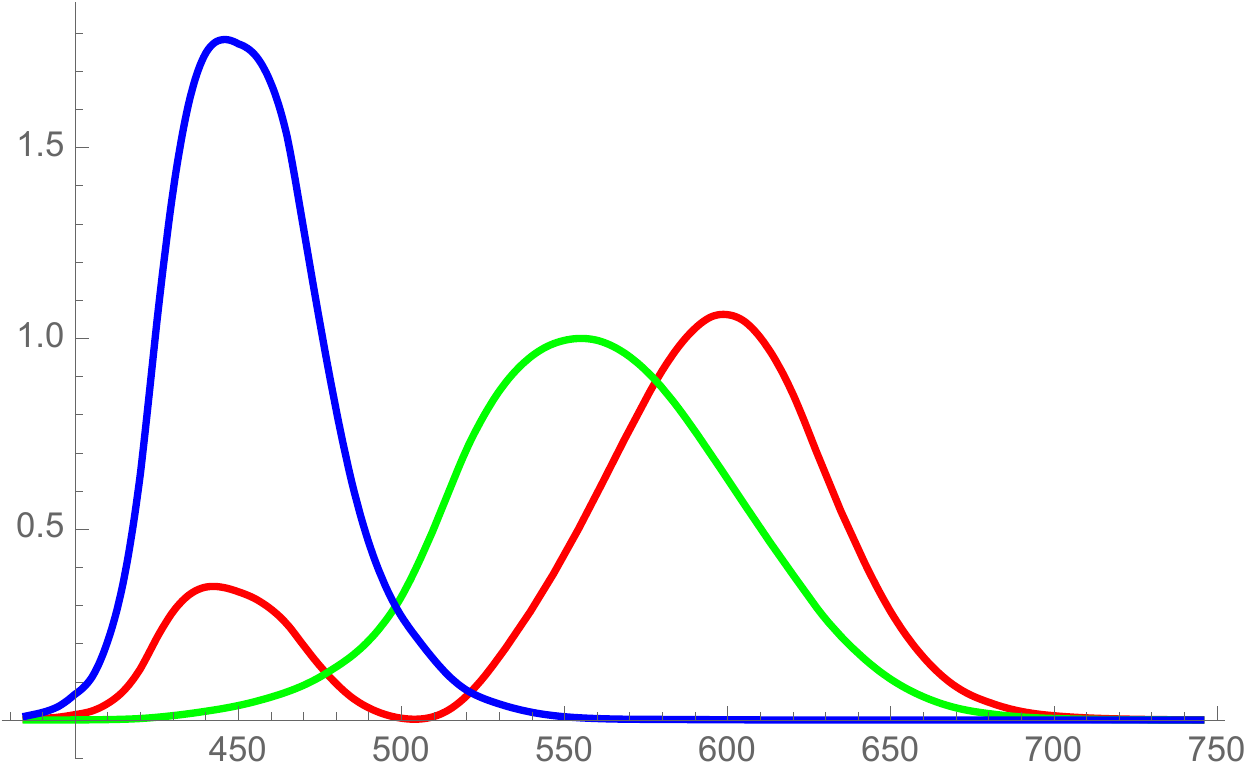}
\captionsetup{justification=raggedright}
\caption{CIE tristimulus matching functions, $x(\lambda)$ (red), $y(\lambda)$ (green), $z(\lambda)$ (blue).}
\label{tristimulus}
\end{center}
\end{figure}

Let us now discuss in more detail how we actually simulated the Doppler shift. The first thing to notice is that human colour sensation involves a huge compression of the full spectrum of wavelengths into excitation rates of only three types of cone cells, sensitive to three different regions of the visible window. So radically different spectral distributions can give rise to the same colour.  This of course also lies at the basis of colour reproduction, with the three types of RGB pixels in digital imagery or the four types of CMYK dots in printing. The RGB values corresponding to a particular spectral power distribution (SPD) $I(\lambda)$ are obtained from the related tristimulus XYZ values, which in turn are computed by integrating the SPD weighed by the CIE matching functions (see \cite{Colour} for more details): 
\be  \{X,Y,Z\}=  \int\!\! d\lambda\,\, \{x(\lambda),y(\lambda),z(\lambda)\}\, I(\lambda)\,. \ee  In figure \ref{tristimulus} we show the precise form of the matching functions. 

\begin{figure}[t]
\begin{center}
\includegraphics[width=0.48\textwidth]{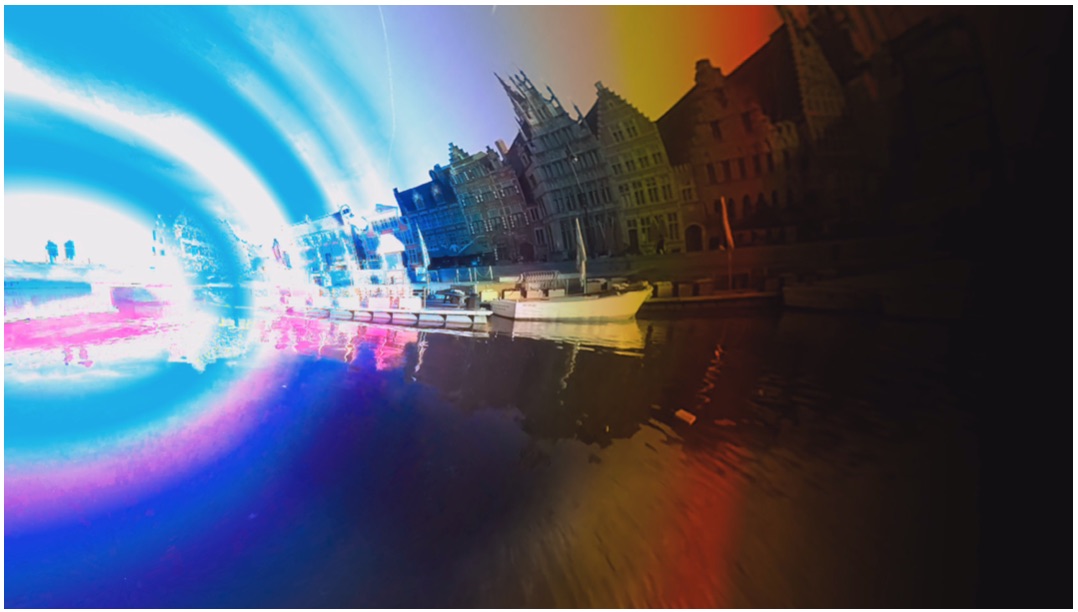}
\captionsetup{justification=raggedright}
\caption{Point of view shot at $v=0.85c$ taking into account the full searchlight effect. (Compare with the corresponding image of the actual movie in figure \ref{figDoppmovie}.) }
\label{figsearchlight}
\end{center}
\end{figure}

To compute the colours of the blue sky on our relativistic boat trip we need the (direction dependent) SPD $I'(\lambda')$ in the boat frame $S$, which follows from the SPD $I(\lambda)$ (figure \ref{figspectrum}) in the quay/sky rest frame $S$ (with $\lambda'=\lambda/D$, see eq. (\ref{doppform})): \be I'(\lambda') d\lambda'=D^4 I(\lambda) d\lambda\,.\label{searchlight}\ee  This is the so called searchlight effect \cite{searchlight}: the light from the blueshifted directions $(D>1)$ gets brighter while the light from the redshifted directions $(D\!<\!1)$ is dimmed.  As discussed above, the colours in front of our boat arise from blueshifted parts of the IR spectrum. Here the full searchlight effect counteracts and even overcompensates the diminished power in the IR tail, resulting in very bright light already for 'moderate velocities' $v\gtrsim 0.7 c$. Together with the darkening of the redshifted regions, this produces images like in figure \ref{figsearchlight}, with oversaturated bright bands in front of the boat on a rather dull dark background. To clearly show all the colour nuances of the Doppler shift we chose to alter this aspect of relativity, by dimming down the searchlight effect for our actual movie, effectively working with a different $D$-dependent pre-factor in (\ref{searchlight}).

Of course the pixels of the input images (e.g. figure \ref{figequi}(a)) only give us RGB values, rather than a full spectrum. To create semi-realistic Doppler shifted images, we assumed realistic underlying SPD's and used these as input for the construction of a transformation at the RGB level. Specifically, we used the computed Doppler shifts of the blue sky spectrum that we discussed above. In addition we also used the Doppler shifts on light reflected from water and from brick, which were computed by taking into account the proper reflection spectra (data obtained from \cite{Spectra}). Our resulting RGB to RGB map then interpolates between the three different cases (sky, brick, water) based on the input RGB-value and the position of the pixel in the original image. 

Some comments are in order here. While our approach should be pretty accurate for the quasi uniform blue sky (barring the aforementioned altered searchlight effect), the colour changes of the other 'materials' in the images are at best a qualitative approximation. We list a few obvious shortcomings: clearly not all buildings or other objects in the movie are made of red brick. Furthermore, certain regions are illuminated by direct sunlight, whereas we assumed incident diffuse sky radiation everywhere. Also, lacking the data, we simply took a flat water reflection spectrum for wavelengths $\lambda > 2250nm$, which starts to have its effect for velocities $v\gtrsim 80 \% c$. Finally, we have also ignored thermal radiation from the different objects. Taking $\lambda_{thermal}\approx 10000 nm$ we can estimate that this would only have a visual effect at ultra-relativistic velocities. 

\vspace{0.2cm}

\noindent {\bf Exercise 9 ($\star\star $): The searchlight effect}

As a last exercise we ask you to verify the searchlight effect (\ref{searchlight}). To this end, show that an incoming photon flux $F=\frac{dN}{dt d\Omega}$ (with $N$ the photon number) in a particular direction on the unit sphere of directions in the reference frame $S$, transforms as $F'=D^3 F$ to an incoming flux $F$ in the transformed direction of the unit sphere in the reference frame $S'$ (see figure \ref{aberration}) . The extra factor $D$ in (\ref{searchlight}) then follows from the energy Doppler shift: $h\nu'=D \,h\nu$.$\rfloor$

\section{Experience}  
In the context of science communication and outreach Captain Einstein has so far been 'tested' on more than 2500 people at different festivals in Belgium and the Netherlands. Our festival setup typically involves a real canoe or kayak (on land), in which the passengers, equipped with a VR set, experience the virtual boat trip. The basic premise of a world with a slow speed of light in combination with the VR experience itself attracts a wide range of people of different backgrounds - young and old, with or without a particular interest in science. The response after viewing the movie is in general very positive - many interesting questions come up, often leading to animated discussions. This helped  shaping our website \cite{Captain} that accompanies the recent online launch of the movie. 

We have also used the movie in the context of the relativity course (third year bachelor physics and astronomy) at Ghent University. All students ($\approx 50$) viewed the movie on a smartphone VR viewer, either at home or in class. This then served as the basis both for a discussion and exercise session in class and a home assignment. During the session in class, for illustrating several aspects of the movie, several VR headsets were passed on by the students, in addition we projected a point-of-view image on a central screen. From the educational perspective, the immersive $360^\circ$ experience of course gives the students a new handle on relativity, notably on the directional dependence of the effects. But in our estimation the main advantage of the movie is actually simply that it gives a fun and concrete setting for studying the basic concepts of special relativity. We refer to \cite{Realtime2} for a study on the positive learning impact of the related Real Time Relativity game.    


\section*{Conclusions}
In this paper we have presented the Captain Einstein project. We discussed how the different effects of special relativity manifest themselves in the movie. Furthermore we have explained how the relativistic light aberration formula (\ref{deform}) lies at the heart of our simulation, as it maps the recorded $360^{\circ}$ images to the corresponding images for a camera moving at relativistic speeds. We also discussed the limitations of this procedure, namely that all objects in the initial reference frame should be at rest. Finally, we described our approach for the semi-realistic simulations of the Doppler shift, assuming physically relevant light spectra behind the RGB pixels in the recorded images. 

As a science communication and outreach tool we evaluate Captain Einstein very positively from our experiences so far. Both the concept and the VR experience itself attract a broad public; and the movie triggers many questions both on relativity and the nature of light and colour. Our website \cite{Captain} was developed to cover some first answers to these questions. 

We have also experimented with Captain Einstein as an educational tool, in the context of a relativity course at the undergraduate university level. Also here our evaluation is positive, it allows for a direct experience of some of the basic special relativistic effects and serves well as the basis for an engaging exercise session. As a final comment, based on our experience we believe VR applications can indeed present a powerful tool for education, but only on top of the 'old school' methods: direct teacher student interaction, blackboards and books (digital or not).    \vspace{0.2cm}

\section*{Acknowledgements} 
 This work was supported by an Odysseus grant from the FWO. We should mention all the collaborators on the production of Captain Einstein, without whom this project would not have been possible: Ingwio D'Hespeel, J\"{u}rgen De Blonde, Nele Vandersickel, Frauke Thies,
Willem Mertens, Wout Standaert, Sarah Verhofstadt, Melissa Somany, Jeffrey Roekens, Gwen Vanhove, Kasper Jordaens, Ignace Saenen, Stefaan Hinneman, Gael Vanhalst and Steven Engels. Also many thanks to the students and friends for guiding the VR festival demos. In addition we thank John De Poorter for his feedback on the use of Captain Einstein at the Artevelde Hogeschool. Finally let us thank Philippe Smet for his valuable comments on an earlier draft.    
  

\begin{thebibliography}{25}%
\makeatletter
\providecommand \@ifxundefined [1]{%
 \@ifx{#1\undefined}
}%
\providecommand \@ifnum [1]{%
 \ifnum #1\expandafter \@firstoftwo
 \else \expandafter \@secondoftwo
 \fi
}%
\providecommand \@ifx [1]{%
 \ifx #1\expandafter \@firstoftwo
 \else \expandafter \@secondoftwo
 \fi
}%
\providecommand \natexlab [1]{#1}%
\providecommand \enquote  [1]{``#1''}%
\providecommand \bibnamefont  [1]{#1}%
\providecommand \bibfnamefont [1]{#1}%
\providecommand \citenamefont [1]{#1}%
\providecommand \href@noop [0]{\@secondoftwo}%
\providecommand \href [0]{\begingroup \@sanitize@url \@href}%
\providecommand \@href[1]{\@@startlink{#1}\@@href}%
\providecommand \@@href[1]{\endgroup#1\@@endlink}%
\providecommand \@sanitize@url [0]{\catcode `\\12\catcode `\$12\catcode
  `\&12\catcode `\#12\catcode `\^12\catcode `\_12\catcode `\%12\relax}%
\providecommand \@@startlink[1]{}%
\providecommand \@@endlink[0]{}%
\providecommand \url  [0]{\begingroup\@sanitize@url \@url }%
\providecommand \@url [1]{\endgroup\@href {#1}{\urlprefix }}%
\providecommand \urlprefix  [0]{URL }%
\providecommand \Eprint [0]{\href }%
\providecommand \doibase [0]{http://dx.doi.org/}%
\providecommand \selectlanguage [0]{\@gobble}%
\providecommand \bibinfo  [0]{\@secondoftwo}%
\providecommand \bibfield  [0]{\@secondoftwo}%
\providecommand \translation [1]{[#1]}%
\providecommand \BibitemOpen [0]{}%
\providecommand \bibitemStop [0]{}%
\providecommand \bibitemNoStop [0]{.\EOS\space}%
\providecommand \EOS [0]{\spacefactor3000\relax}%
\providecommand \BibitemShut  [1]{\csname bibitem#1\endcsname}%
\let\auto@bib@innerbib\@empty
\bibitem [{Cap()}]{Captain}%
  \BibitemOpen
  \href@noop {} {\bibinfo  {journal}
  {\href{http://www.captaineinstein.org}{www.captaineinstein.org}}\
  }\BibitemShut {NoStop}%
\bibitem [{\citenamefont {Gamow}(1972)}]{Gamow}%
  \BibitemOpen
\bibfield  {journal} {  }\bibfield  {author} {\bibinfo {author} {\bibfnamefont
  {George}\ \bibnamefont {Gamow}},\ }\bibfield  {title} {\enquote {\bibinfo
  {title} {Mr. tompkins},}\ }\href@noop {} {\bibfield  {journal} {\bibinfo
  {journal} {Cambridge, University Press}\ } (\bibinfo {year}
  {1972})}\BibitemShut {NoStop}%
\bibitem [{Tub()}]{Tubingen}%
  \BibitemOpen
  \href@noop {} {\bibinfo  {journal}
  {\href{https://www.spacetimetravel.org/tuebingen/tuebingen.html}{https://www.spacetimetravel.org/tuebingen/}}\
  }\BibitemShut {NoStop}%
\bibitem [{\citenamefont {Kraus}\ and\ \citenamefont
  {Borchers}(2005)}]{Tubingen2}%
  \BibitemOpen
\bibfield  {journal} {  }\bibfield  {author} {\bibinfo {author} {\bibfnamefont
  {U.}~\bibnamefont {Kraus}}\ and\ \bibinfo {author} {\bibfnamefont {M.~P.}\
  \bibnamefont {Borchers}},\ }\bibfield  {title} {\enquote {\bibinfo {title}
  {Fast lichtschnell durch die stadt: Visualisierung relativistischer
  effekte},}\ }\href@noop {} {\bibfield  {journal} {\bibinfo  {journal} {Physik
  in unserer Zeit}\ }\textbf {\bibinfo {volume} {36}},\ \bibinfo {pages}
  {64--69} (\bibinfo {year} {2005})}\BibitemShut {NoStop}%
\bibitem [{RTR()}]{RTR}%
  \BibitemOpen
  \href@noop {} {\bibinfo  {journal}
  {\href{http://www.anu.edu.au/Physics/Savage/RTR}{www.anu.edu.au/Physics/Savage/RTR}}\
  }\BibitemShut {NoStop}%
\bibitem [{\citenamefont {{Savage}}\ \emph {et~al.}(2007)\citenamefont
  {{Savage}}, \citenamefont {{Searle}},\ and\ \citenamefont
  {{McCalman}}}]{Realtime}%
  \BibitemOpen
\bibfield  {journal} {  }\bibfield  {author} {\bibinfo {author} {\bibfnamefont
  {C.~M.}\ \bibnamefont {{Savage}}}, \bibinfo {author} {\bibfnamefont
  {A.}~\bibnamefont {{Searle}}}, \ and\ \bibinfo {author} {\bibfnamefont
  {L.}~\bibnamefont {{McCalman}}},\ }\bibfield  {title} {\enquote {\bibinfo
  {title} {{Real Time Relativity: Exploratory learning of special
  relativity}},}\ }\href {\doibase 10.1119/1.2744048} {\bibfield  {journal}
  {\bibinfo  {journal} {American Journal of Physics}\ }\textbf {\bibinfo
  {volume} {75}},\ \bibinfo {pages} {791--798} (\bibinfo {year}
  {2007})}\BibitemShut {NoStop}%
\bibitem [{\citenamefont {{McGrath}}\ \emph {et~al.}(2010)\citenamefont
  {{McGrath}}, \citenamefont {{Wegener}}, \citenamefont {{McIntyre}},
  \citenamefont {{Savage}},\ and\ \citenamefont {{Williamson}}}]{Realtime2}%
  \BibitemOpen
  \bibfield  {author} {\bibinfo {author} {\bibfnamefont {D.}~\bibnamefont
  {{McGrath}}}, \bibinfo {author} {\bibfnamefont {M.}~\bibnamefont
  {{Wegener}}}, \bibinfo {author} {\bibfnamefont {T.~J.}\ \bibnamefont
  {{McIntyre}}}, \bibinfo {author} {\bibfnamefont {C.}~\bibnamefont
  {{Savage}}}, \ and\ \bibinfo {author} {\bibfnamefont {M.}~\bibnamefont
  {{Williamson}}},\ }\bibfield  {title} {\enquote {\bibinfo {title} {{Student
  experiences of virtual reality: A case study in learning special
  relativity}},}\ }\href {\doibase 10.1119/1.3431565} {\bibfield  {journal}
  {\bibinfo  {journal} {American Journal of Physics}\ }\textbf {\bibinfo
  {volume} {78}},\ \bibinfo {pages} {862--868} (\bibinfo {year}
  {2010})}\BibitemShut {NoStop}%
\bibitem [{mit()}]{mit}%
  \BibitemOpen
  \href@noop {} {\bibinfo  {journal}
  {\href{http://gamelab.mit.edu/games/a-slower-speed-of-light/}
  {http://gamelab.mit.edu/games/a-slower-speed-of-light/}}\ }\BibitemShut
  {NoStop}%
\bibitem [{\citenamefont {{Kortemeyer}}\ \emph {et~al.}(2013)\citenamefont
  {{Kortemeyer}}, \citenamefont {{Fish}}, \citenamefont {{Hacker}},
  \citenamefont {{Kienle}}, \citenamefont {{Kobylarek}}, \citenamefont
  {{Sigler}}, \citenamefont {{Wierenga}}, \citenamefont {{Cheu}}, \citenamefont
  {{Kim}}, \citenamefont {{Sherin}}, \citenamefont {{Sidhu}},\ and\
  \citenamefont {{Tan}}}]{mit2}%
  \BibitemOpen
\bibfield  {journal} {  }\bibfield  {author} {\bibinfo {author} {\bibfnamefont
  {G.}~\bibnamefont {{Kortemeyer}}}, \bibinfo {author} {\bibfnamefont
  {J.}~\bibnamefont {{Fish}}}, \bibinfo {author} {\bibfnamefont
  {J.}~\bibnamefont {{Hacker}}}, \bibinfo {author} {\bibfnamefont
  {J.}~\bibnamefont {{Kienle}}}, \bibinfo {author} {\bibfnamefont
  {A.}~\bibnamefont {{Kobylarek}}}, \bibinfo {author} {\bibfnamefont
  {M.}~\bibnamefont {{Sigler}}}, \bibinfo {author} {\bibfnamefont
  {B.}~\bibnamefont {{Wierenga}}}, \bibinfo {author} {\bibfnamefont
  {R.}~\bibnamefont {{Cheu}}}, \bibinfo {author} {\bibfnamefont
  {E.}~\bibnamefont {{Kim}}}, \bibinfo {author} {\bibfnamefont
  {Z.}~\bibnamefont {{Sherin}}}, \bibinfo {author} {\bibfnamefont
  {S.}~\bibnamefont {{Sidhu}}}, \ and\ \bibinfo {author} {\bibfnamefont
  {P.}~\bibnamefont {{Tan}}},\ }\bibfield  {title} {\enquote {\bibinfo {title}
  {{Seeing and Experiencing Relativity -- A New Tool for Teaching?}}}\ }\href
  {\doibase 10.1119/1.4824935} {\bibfield  {journal} {\bibinfo  {journal} {The
  Physics Teacher}\ }\textbf {\bibinfo {volume} {51}},\ \bibinfo {pages}
  {460--461} (\bibinfo {year} {2013})}\BibitemShut {NoStop}%
\bibitem [{\citenamefont {{Sherin}}\ \emph {et~al.}(2016)\citenamefont
  {{Sherin}}, \citenamefont {{Cheu}}, \citenamefont {{Tan}},\ and\
  \citenamefont {{Kortemeyer}}}]{mit3}%
  \BibitemOpen
  \bibfield  {author} {\bibinfo {author} {\bibfnamefont {Z.~W.}\ \bibnamefont
  {{Sherin}}}, \bibinfo {author} {\bibfnamefont {R.}~\bibnamefont {{Cheu}}},
  \bibinfo {author} {\bibfnamefont {P.}~\bibnamefont {{Tan}}}, \ and\ \bibinfo
  {author} {\bibfnamefont {G.}~\bibnamefont {{Kortemeyer}}},\ }\bibfield
  {title} {\enquote {\bibinfo {title} {{Visualizing relativity: The
  OpenRelativity project}},}\ }\href {\doibase 10.1119/1.4938057} {\bibfield
  {journal} {\bibinfo  {journal} {American Journal of Physics}\ }\textbf
  {\bibinfo {volume} {84}},\ \bibinfo {pages} {369--374} (\bibinfo {year}
  {2016})}\BibitemShut {NoStop}%
\bibitem [{\citenamefont {Weiskopf}(2001)}]{Weiskopf}%
  \BibitemOpen
  \bibfield  {author} {\bibinfo {author} {\bibfnamefont {D.}~\bibnamefont
  {Weiskopf}},\ }\bibfield  {title} {\enquote {\bibinfo {title} {Visualization
  of four-dimensional spacetimes},}\ }\href@noop {} {\bibfield  {journal}
  {\bibinfo  {journal} {PhD-dissertation, University of T\"{u}bingen, \href{
  https://publikationen.uni-tuebingen.de/xmlui/handle/10900/48159}{
  http://nbn-resolving.de/urn: nbn:de:bsz:21-opus-2400}}\ } (\bibinfo {year}
  {2001})}\BibitemShut {NoStop}%
\bibitem [{\citenamefont {{Sherin}}\ \emph {et~al.}(2017)\citenamefont
  {{Sherin}}, \citenamefont {{Tan}}, \citenamefont {{Fairweather}},\ and\
  \citenamefont {{Kortemeyer}}}]{dome}%
  \BibitemOpen
  \bibfield  {author} {\bibinfo {author} {\bibfnamefont {Z.}~\bibnamefont
  {{Sherin}}}, \bibinfo {author} {\bibfnamefont {P.}~\bibnamefont {{Tan}}},
  \bibinfo {author} {\bibfnamefont {H.}~\bibnamefont {{Fairweather}}}, \ and\
  \bibinfo {author} {\bibfnamefont {G.}~\bibnamefont {{Kortemeyer}}},\
  }\bibfield  {title} {\enquote {\bibinfo {title} {{``Einstein's Playground'':
  An Interactive Planetarium Show on Special Relativity}},}\ }\href {\doibase
  10.1119/1.5011832} {\bibfield  {journal} {\bibinfo  {journal} {The Physics
  Teacher}\ }\textbf {\bibinfo {volume} {55}},\ \bibinfo {pages} {550--554}
  (\bibinfo {year} {2017})}\BibitemShut {NoStop}%
\bibitem [{\citenamefont {Younsi}()}]{VRBH1}%
  \BibitemOpen
  \bibfield  {author} {\bibinfo {author} {\bibfnamefont {Ziri}\ \bibnamefont
  {Younsi}},\ }\bibfield  {title} {\enquote {\bibinfo {title} {Falling into a
  black hole},}\ }\href@noop {} {\bibinfo  {journal}
  {\href{http://ziriyounsi.com/videos/}{http://ziriyounsi.com/videos/}}\
  }\BibitemShut {NoStop}%
\bibitem [{VRB()}]{VRBH2}%
  \BibitemOpen
\bibfield  {journal} {  }\bibfield  {title} {\enquote {\bibinfo {title} {Plunge
  into a (virtual reality) black hole},}\ }\href@noop {} {\bibinfo  {journal}
  {\href{https://www.quantamagazine.org/plunge-into-a-virtual-reality-black-hole-20171204/}{https://www.quantamagazine.org/}}\
  }\BibitemShut {NoStop}%
\bibitem [{\citenamefont {{Lampa}}(1924)}]{Lampa}%
  \BibitemOpen
\bibfield  {journal} {  }\bibfield  {author} {\bibinfo {author} {\bibfnamefont
  {A.}~\bibnamefont {{Lampa}}},\ }\bibfield  {title} {\enquote {\bibinfo
  {title} {{Wie erscheint nach der Relativit{\"a}tstheorie ein bewegter Stab
  einem ruhenden Beobachter?}}}\ }\href {\doibase 10.1007/BF01328021}
  {\bibfield  {journal} {\bibinfo  {journal} {Zeitschrift fur Physik}\ }\textbf
  {\bibinfo {volume} {27}},\ \bibinfo {pages} {138--148} (\bibinfo {year}
  {1924})}\BibitemShut {NoStop}%
\bibitem [{\citenamefont {{Penrose}}(1959)}]{Penrose}%
  \BibitemOpen
  \bibfield  {author} {\bibinfo {author} {\bibfnamefont {R.}~\bibnamefont
  {{Penrose}}},\ }\bibfield  {title} {\enquote {\bibinfo {title} {{The apparent
  shape of a relativistically moving sphere}},}\ }\href {\doibase
  10.1017/S0305004100033776} {\bibfield  {journal} {\bibinfo  {journal}
  {Proceedings of the Cambridge Philosophical Society}\ }\textbf {\bibinfo
  {volume} {55}},\ \bibinfo {pages} {137} (\bibinfo {year} {1959})}\BibitemShut
  {NoStop}%
\bibitem [{\citenamefont {{Terrell}}(1959)}]{Terell}%
  \BibitemOpen
  \bibfield  {author} {\bibinfo {author} {\bibfnamefont {J.}~\bibnamefont
  {{Terrell}}},\ }\bibfield  {title} {\enquote {\bibinfo {title} {{Invisibility
  of the Lorentz Contraction}},}\ }\href {\doibase 10.1103/PhysRev.116.1041}
  {\bibfield  {journal} {\bibinfo  {journal} {Physical Review}\ }\textbf
  {\bibinfo {volume} {116}},\ \bibinfo {pages} {1041--1045} (\bibinfo {year}
  {1959})}\BibitemShut {NoStop}%
\bibitem [{\citenamefont {{Einstein}}(1905)}]{Einstein}%
  \BibitemOpen
  \bibfield  {author} {\bibinfo {author} {\bibfnamefont {A.}~\bibnamefont
  {{Einstein}}},\ }\bibfield  {title} {\enquote {\bibinfo {title} {{Zur
  Elektrodynamik bewegter K{\"o}rper}},}\ }\href {\doibase
  10.1002/andp.19053221004} {\bibfield  {journal} {\bibinfo  {journal} {Annalen
  der Physik}\ }\textbf {\bibinfo {volume} {322}},\ \bibinfo {pages} {891--921}
  (\bibinfo {year} {1905})}\BibitemShut {NoStop}%
\bibitem [{\citenamefont {Hogg}()}]{hogg}%
  \BibitemOpen
  \bibfield  {author} {\bibinfo {author} {\bibfnamefont {David~W.}\
  \bibnamefont {Hogg}},\ }\bibfield  {title} {\enquote {\bibinfo {title}
  {{Special Relativity}},}\ }\href@noop {} {\bibinfo  {journal}
  {\href{http://cosmo.nyu.edu/hogg/sr/sr.pdf}{http://cosmo.nyu.edu/hogg/sr/sr.pdf}}\
  }\BibitemShut {NoStop}%
\bibitem [{pho()}]{photopic}%
  \BibitemOpen
\bibfield  {journal} {  }\href@noop {} {\bibinfo  {journal}
  {\href{https://light-measurement.com/spectral-sensitivity-of-eye/}{https://light-measurement.com/spectral-sensitivity-of-eye/}}\
  }\BibitemShut {NoStop}%
\bibitem [{\citenamefont {Laboratory}()}]{Smarts}%
  \BibitemOpen
\bibfield  {journal} {  }\bibfield  {author} {\bibinfo {author} {\bibfnamefont
  {National Renewable~Energy}\ \bibnamefont {Laboratory}},\ }\bibfield  {title}
  {\enquote {\bibinfo {title} {Smarts: Simple model of the radiative transfer
  of sunshine (example 1:ussa astm iso old std)},}\ }\href@noop {} {\bibinfo
  {journal}
  {\href{https://www.nrel.gov/rredc/smarts/}{https://www.nrel.gov/rredc/smarts/}}\
  }\BibitemShut {NoStop}%
\bibitem [{\citenamefont {J.C.~Hafele}(1972)}]{hafelekeating}%
  \BibitemOpen
\bibfield  {journal} {  }\bibfield  {author} {\bibinfo {author} {\bibfnamefont
  {R.~E.~Keating}\ \bibnamefont {J.C.~Hafele}},\ }\bibfield  {title} {\enquote
  {\bibinfo {title} {Around-the-world atomic clocks: Predicted relativistic
  time gains},}\ }\href {\doibase https://doi.org/10.1126/science.177.4044.166}
  {\bibfield  {journal} {\bibinfo  {journal} {Science}\ }\textbf {\bibinfo
  {volume} {177}},\ \bibinfo {pages} {166--168} (\bibinfo {year}
  {1972})}\BibitemShut {NoStop}%
\bibitem [{\citenamefont {Pointer}(2011)}]{Colour}%
  \BibitemOpen
  \bibfield  {author} {\bibinfo {author} {\bibfnamefont {R.~W. G. Hunt M.~R.}\
  \bibnamefont {Pointer}},\ }\bibfield  {title} {\enquote {\bibinfo {title}
  {{Measuring Colour}},}\ }\href@noop {} {\bibfield  {journal} {\bibinfo
  {journal} {Wiley}\ } (\bibinfo {year} {2011})}\BibitemShut {NoStop}%
\bibitem [{\citenamefont {{McKinley}}(1979)}]{searchlight}%
  \BibitemOpen
  \bibfield  {author} {\bibinfo {author} {\bibfnamefont {J.~M.}\ \bibnamefont
  {{McKinley}}},\ }\bibfield  {title} {\enquote {\bibinfo {title}
  {{Relativistic transformations of light power}},}\ }\href {\doibase
  10.1119/1.11762} {\bibfield  {journal} {\bibinfo  {journal} {American Journal
  of Physics}\ }\textbf {\bibinfo {volume} {47}},\ \bibinfo {pages} {602--605}
  (\bibinfo {year} {1979})}\BibitemShut {NoStop}%
\bibitem [{\citenamefont {Library}()}]{Spectra}%
  \BibitemOpen
  \bibfield  {author} {\bibinfo {author} {\bibfnamefont {USGS~Spectral}\
  \bibnamefont {Library}},\ }\href@noop {} {\bibinfo  {journal}
  {\href{https://speclab.cr.usgs.gov/spectral-lib.html}{https://speclab.cr.usgs.gov/spectral-lib.html}}\
  }\BibitemShut {NoStop}%
\end{thebibliography}

%

\section*{Exercise solutions}

\noindent{\bf Exercise 1}

(a) The length $l'$ between the two bridges as measured in the boat frame is contracted: $l'=l/\gamma(v)$. This gives us:  \be \Delta t'=\frac{l}{v\gamma (v)}\,. \ee So we have that for $v\rightarrow c$, the boat crosses the distance between the two bridges at a time $\Delta t'\rightarrow 0$. Even in a world with maximal velocity $c=20km/h$ we can in principle get everywhere as fast as we want, simply because all distances are length contracted to zero! 

(b) From the perspective of the quay rest frame the distance between the bridges is of course independent of the boat velocity $v$, so we have $\Delta t=l/v$ for the measured boat travel time. But if we now take into account the time dilation ("\emph{...time is slowing down, as we speed up}") for the moving clock on board the boat, we recover the result of (a): $\Delta t'=\Delta t/\gamma(v)$. 

(c) In the real world  $v_{b}/c\approx 0$, which means that we can interpret the frames in the original footage as a series of pictures at different positions in the quay rest frame. The speed-up of the movie should correspond to e.g. the number of houses (or bridges) that pass by per unit time, not the distance travelled per unit time. In light of (a) (or (b)) it is clear then that we should speed up the movie by a factor:  $\gamma(v)\frac{v}{v_{b}}.$  
\vspace{0.2cm}

\noindent{\bf Exercise 2}

Let us compute the g-force corresponding to an acceleration $a=\frac{dv}{dt'}$ at $t'=t'_0$. At that time the infinitesimal velocity change reads: $v(t_0'+dt')=v_0+ a_0\, dt'$, with $v_0\equiv v(t_0')$ and $a_0\equiv\frac{dv}{dt'}|_{t'=t_0}$. To obtain the corresponding g-force we boost to the instantaneous inertial boat rest frame $S(v_0)$. In this inertial rest frame we can use Newtonian physics to equate the acceleration to the experienced g-force. The velocity addition formula, going from $S=S(0)$ to $\bar{S}=S(v_0)$, reads:  
\be \bar{v}=\frac{v-v_0}{1-\frac{v_0 v}{c^2}}\,.\ee Plugging in $v(t_0'+dt')=v_0+ a_0\, dt'$ then gives us: 
\be \bar{v}(t_0'+dt')=\frac{a_0 dt'}{1-\frac{v_0^2}{c^2}},\ee
which finally allows us to read off the g-force $g$, corresponding to an acceleration $a=\frac{dv}{dt'}$ at velocity $v$: \be g=\frac{a}{1-v^2/c^2}=\gamma^2(u)a\,.\ee

For a constant acceleration $a=0.003\,G$, assuming that we pass out at $10\,G$, we then find a critical velocity $v_{crit}=19.997 km/h$. \vspace{0.2cm}

\noindent{\bf Exercise 3}

The top of the tower was positioned more to the left than the bottom of the tower when the respective photons left, leading to the distorted picture in figure \ref{figtowers}. \vspace{0.2cm}

\noindent{\bf Exercise 4}

When the centre of the fence appears in the centre of our viewfinder, aiming at an angle $\theta'=\pi/2$, and assuming that the fence lies parallel to the velocity direction, we know that the travelled distance of the photons from both endpoints to our camera is the same. The picture therefore shows both endpoints at the same time $t'$ in our reference frame $S'$. And we can therefore indeed use standard Euclidean geometry in our 3D-space to relate the length $l'$ of the fence, to the opening angle $\alpha$ and the distance $y$ to the boat: $l'=2 y \tan(\alpha/2).$  

For the house at the left, the photons coming from the left endpoint travelled over a longer distance than those coming from the right endpoint. The left endpoint is therefore shown at an earlier time in the past than the right endpoint: $t'_L<t'_R$. So at $t'_L$ the house was positioned more to the left than at $t'_R$. This elongates the apparent shape in the picture, counteracting the length contraction. For the van at the left, we now have that $t'_L>t'_R$, producing an extra apparent contraction in addition to the length contraction. \vspace{0.2cm}

\noindent{\bf Exercise 5}

First of all, notice that in the quay rest frame we do not have to worry about the vision delay effect as the endpoints of the fence do not move. From figure \ref{figabcon} we then find: \be l=y\left(\cot\theta_1-\cot \theta_2\right)\,.\label{lcot}\ee We can then use the aberration formula (\ref{deform}) to get an expression in terms of the angles $\theta'$ in the boat rest frame. First, we can easily invert (\ref{deform}), by substituting $v \rightarrow -v$: \be \cos\theta=\frac{\cos\theta'-\frac{v}{c}}{1-\frac{v}{c}\cos\theta'}\,,\ee after which we can also solve for $\sin\theta$:
\be \sin\theta=\sqrt{1-\cos^2\theta}=\frac{\sin \theta' \sqrt{1-\frac{v^2}{c^2}}}{1-\frac{v}{c}\cos\theta'}\,,\ee
arriving at: \be \cot\theta=\frac{\cos \theta}{\sin \theta}=\gamma(v)\frac{\cos\theta'-\frac{v}{c}}{\sin\theta'}\,.\ee
With $\theta_1'=\pi/2-\alpha/2$ and $\theta_2'=\pi/2+\alpha/2$, eq. (\ref{lcot}) then gives us:
\be l=2\gamma(v)\tan(\frac{\alpha}{2})=\gamma(v)l' \,,\ee which is precisely the length contraction formula.\vspace{0.2cm}

\noindent{\bf Exercise 6}

\begin{figure}[t]
\begin{center}
\includegraphics[width=0.5\textwidth]{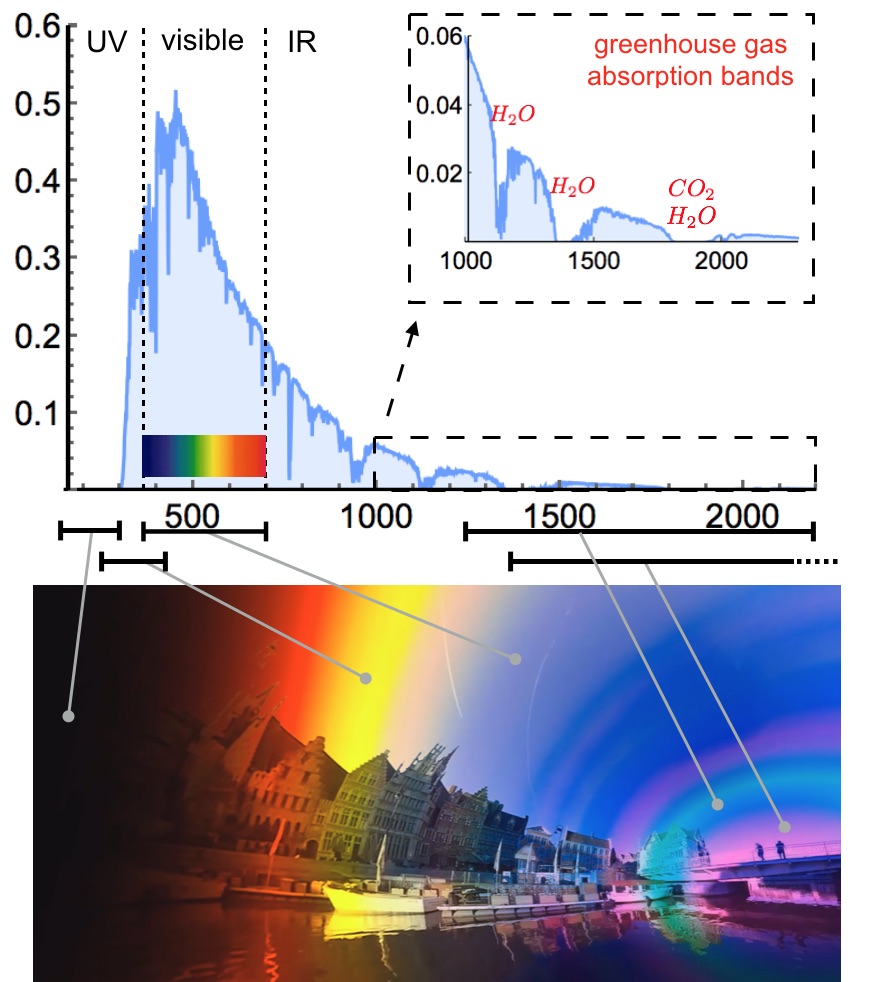}
\captionsetup{justification=raggedright}
\caption{For directions $\theta'=\{105^{\circ},80^{\circ},56^{\circ},12^{\circ},3^{\circ}\}$ (and for $v=0.85 c$), the different intervals that are Doppler shifted into the visible window $[390nm,700nm]$, producing the different colour effects on our relativistic boat trip.}
\label{figdoppintervals}
\end{center}
\end{figure} 

To understand the origin of the different colours in figure \ref{figdopp}(b) or figure \ref{figDoppmovie} you have to look at the particular wavelength intervals that are Doppler shifted (by eq. (\ref{doppform})) into the visible window for the different viewing directions $\theta'$, see figure \ref{figdoppintervals}. In this way you can understand that the yellow and red colours for orthogonal viewing angles $\theta'\approx 90^{\circ}$ originate from the UV band $[300nm,400nm]$ in the blue sky spectrum, which for increasing angles $\theta'\gtrsim 80^{\circ}$ gets redshifted further and further into the low frequency part of the visible window $[580nm,700nm]$. This up to the point that the UV band is shifted completely beyond the visible window, resulting in a darkness for directions $\theta'\gtrsim100^{\circ}$. So the reason for the  \emph{dark behind you} is simply the absence of UV radiation in the blue sky spectrum for wavelengths $\lambda\lesssim300nm$.

Similarly the distinct blue and purple bands for small angles $\theta'$ arise from the IR part of the spectrum that is now blueshifted $(D>1)$ into the visble window. Were it not for the greenhouse absorption bands, the IR tail would be smooth, resulting in a rather uniform colour, independent of the precise viewing direction $\theta'$. However, the absorption bands introduce sharp features, producing distinct colours depending on the particular position of the Doppler shifted bands in the visible spectrum. (See the end of section III for more details on the relation between spectra and colours.)\vspace{0.2cm}

\noindent{\bf Exercise 7}

\begin{figure}[t]
\begin{center}
\includegraphics[width=0.48\textwidth]{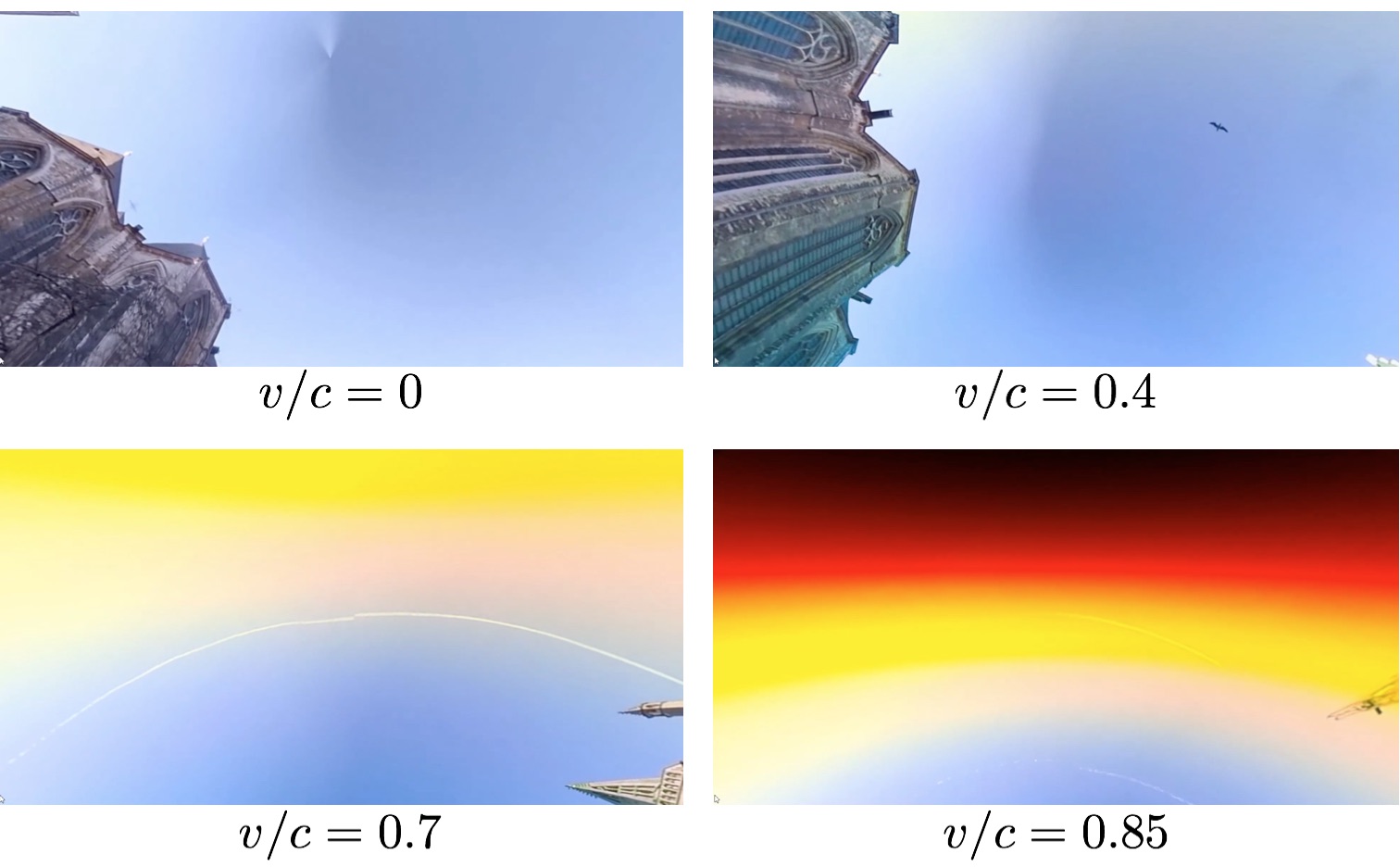}
\captionsetup{justification=raggedright}
\caption{The transverse Doppler effect: point of view shots at different speeds, always looking straight up.}
\label{figtransvers}
\end{center}
\end{figure}

The transverse Doppler effect can be simply observed by looking straight up during the boat trip (see figure \ref{figtransvers}). Notice that the $v/c=0.4$ image does not differ much from the $v=0$ view: the transverse Doppler effect is a truly relativistic order $v^2/c^2$ effect, manifesting itself rather late during the boat trip. (And it is equivalent to the time dilation, as we discussed in the exercise question).\vspace{0.2cm} 

\noindent{\bf Exercise 8}

(a) The pedestrian's velocity $u'$ in the quay frame follows immediately from the velocity addition formula: \be u'=\frac{u+v}{1+\frac{vu}{c^2}}\,.\ee

(b) The Doppler formula applies for any periodic signal that is transmitted with light speed (identifying  $\lambda=c/f$ with $f$ the signal frequency). By using the Doppler formula to go from the pedestrian's reference frame $S_0$ to the boat frame $S'$, we then immediately find \be f'=f_0\frac{\sqrt{1-\frac{{u'}^2}{c^2}}}{1-\frac{u'}{c}\cos\theta'}\,,\ee
for the visually observed walking pace $f'$. Solving $f'=f_0$ we then find a critical angle $\theta_c$  \be \theta_c=\arccos \left(\frac{c}{u'}-\sqrt{\frac{c^2}{{u'}^2}-1}\right) <\pi/2\,. \ee  For viewing directions $\theta'<\theta_c$ we would see the properly visualized pedestrian in fast motion, while for  $\theta'>\theta_c$ we would see her in slow motion. Notice that in orthogonal directions $\theta'=\pi/2$ we would see a walking pace $f'$ slowed down by a factor $\sqrt{1-{u'}^2/c^2}$. This is of course again the transverse Doppler effect, a visual manifestation of the time dilation, in this case showing us directly the slowdown of the pedestrian's eigentime with respect to the boat frame time $t$.  \vspace{0.2cm}

\noindent{\bf Exercise 9}

Similar to the previous exercise we can apply the Doppler formula also on the photon flux 'frequency' $f=\frac{dN}{dt}$:
\be \frac{dN'}{dt'}= D \frac{dN}{dt}\,.\ee While for the surface element $d\Omega=d\!\cos\theta d\varphi$, we have from the inverted aberration formula (\ref{deform}) (replacing $v\rightarrow -v$, $\theta\leftrightarrow \theta'$): \be d\Omega=d\!\cos\theta d\varphi=\frac{1-v^2/c^2}{(1-\frac{v}{c}\cos\theta')^2}d\!\cos\theta'd\varphi=D^2 d\Omega'\,.\ee

Together this amounts to: \be F'=\frac{dN'}{dt'd\Omega'}=D^3 \frac{dN}{dt d\Omega}=D^3 F\,. \ee

\end{document}